\RequirePackage{fix-cm}
\documentclass[twocolumn]{svjour3}          
\smartqed  
\usepackage{graphicx}
\usepackage{graphicx}				
\usepackage{amssymb}
\usepackage{amsmath}
\usepackage{cite}
\usepackage{color}
\usepackage{float}
\usepackage{url}
\usepackage{setspace}
\usepackage[colorinlistoftodos]{todonotes}
\usepackage{stfloats}
\begin{document}

\title{The role of mechanics in the growth and homeostasis of the intestinal crypt}


\author{A.A. Almet$^{1, 2 ,3}$ \and H.M. Byrne$^1$ \and P.K. Maini$^1$ \and D.E. Moulton$^1$}
\date{}							

\institute{
	$^1$ Wolfson Centre for Mathematical Biology, Mathematical Institute, University of Oxford, Andrew Wiles Building, Radcliffe Observatory Quarter, Woodstock Road, Oxford OX2 6GG, United Kingdom; $^2$ NSF-Simons Center for Multiscale Cell Fate Research, University of California, Irvine, Irvine, California; $^3$ Department of Mathematics, University of California, Irvine, Irvine, California
}

\date{}

\maketitle

\begin{abstract}
We present a mechanical model of tissue homeostasis that is specialised to the intestinal crypt. Growth and deformation of the crypt, idealised as a line of cells on a substrate, are modelled using morphoelastic rod theory. Alternating between Lagrangian and Eulerian mechanical descriptions enables us precisely to characterise the dynamic nature of tissue homeostasis, whereby the proliferative structure and morphology are static in the Eulerian frame, but there is active migration of Lagrangian material points out of the crypt. Assuming mechanochemical growth, we identify the necessary conditions for homeostasis, reducing the full, time-dependent system to a static boundary value problem characterising a spatially heterogeneous ``treadmilling'' state. We extract essential features of crypt homeostasis, such as the morphology, the proliferative structure, the migration velocity, and the sloughing rate. We also derive closed-form solutions for growth and sloughing dynamics in homeostasis, and show that mechanochemical growth is sufficient to generate the observed proliferative structure of the crypt. Key to this is the concept of \emph{threshold-dependent} mechanical feedback, that regulates an established Wnt signal for biochemical growth. Numerical solutions demonstrate the importance of crypt morphology on homeostatic growth, migration, and sloughing, and highlight the value of this framework as a foundation for studying the role of mechanics in homeostasis.
\end{abstract}

\section{Introduction}
The crypts of Liehberk\"{u}hn are a canonical example of biochemistry and biomechanics combining to maintain tissue homeostasis within a highly-deformed morphology. These test-tube-shaped invaginations renew and maintain a protective epithelial layer, called the intestinal epithelium, for the small intestine and colon. In the context of disease, colonic cancer originates in the crypts \cite{Humphries2008}, while during inflammation, crypts facilitate rapid regeneration of the epithelium \cite{seno2009efficient}. Therefore, proper crypt function is crucial to a healthy gut. Deciphering the numerous genetic and biochemical signalling pathways governing crypt homeostasis has been the focus of a significant amount of research. Mathematical and computational modelling has been particularly useful in providing insight. However, many aspects of crypt morphogenesis and homeostasis are still not well understood. 

\begin{figure*}[t!]
	\centering
	\includegraphics[width=\textwidth]{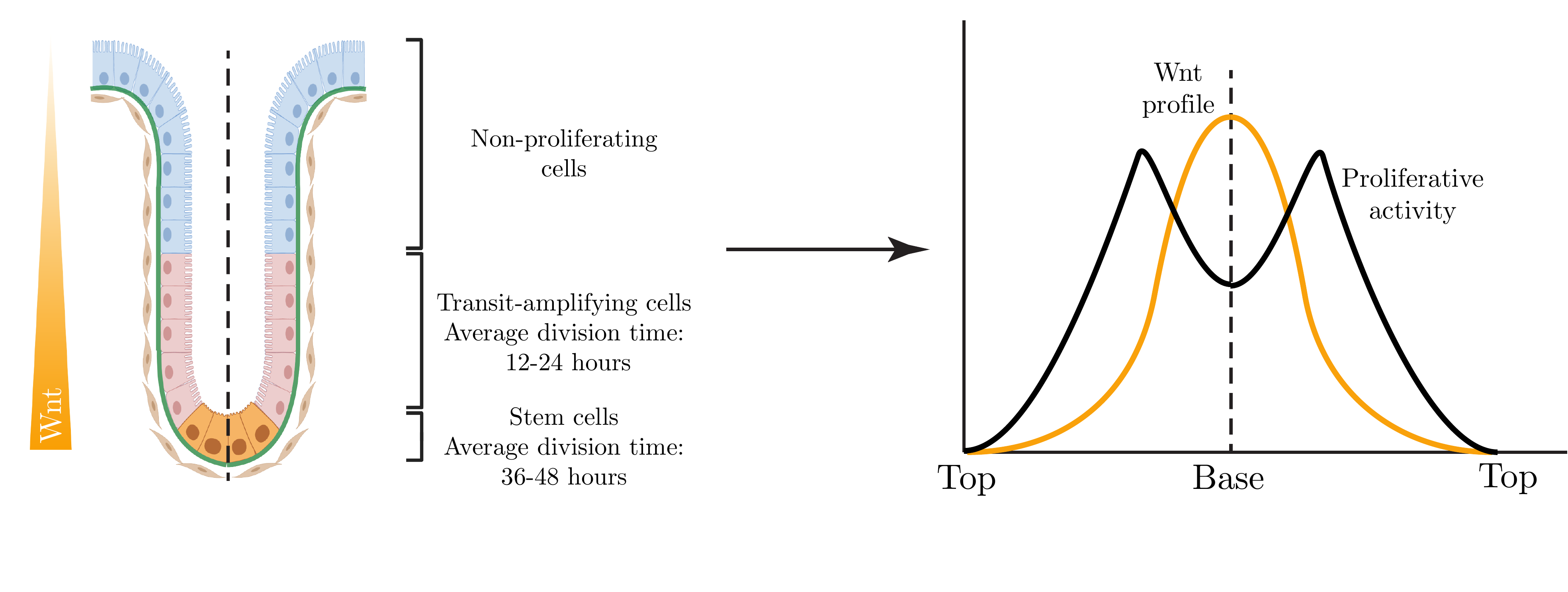}
	\vspace{-1cm}
	\caption{\textbf{The internal proliferative structure of the crypt.} The proliferative structure is bimodal as a function of position along the crypt, i.e. maximal between the crypt base and crypt top. However, Wnt signalling, thought to be the primary governor of proliferation, is unimodal, i.e. maximal at the base.}
	\label{fig:growthstructurecryptdiagram}
\end{figure*}

One aspect of uncertainty concerns the unique and robust proliferative structure of the crypt. In the base of the crypt resides a pool of stem cells, which produce progenitors that migrate upwards. Transit-amplifying cells are the first progenitor cell type to emerge, proliferating rapidly for a fixed number of divisions as they migrate from the crypt base. Transit-amplifying cells differentiate into non-proliferating specialised cells, which reside at the top of the crypt. Despite the robustness of this hierarchical structure, it is not fully understood how it emerges. Wnt signalling is known to be a primary driver of proliferation within the crypt \cite{clevers2006wnt} and forms a decreasing spatial profile from the crypt base to the top \cite{gaspar2004apc}. If proliferation within the crypt were driven solely by Wnt, then proliferative activity would be highest in the base with a monotonically decreasing profile moving towards the top; viewed as a function of arc length along a single crypt from top to top, we would observe a ``unimodal'' form of growth, peaking in the middle (the base). However, proliferative activity is concentrated in the transit-amplifying cell region, creating instead a ``bimodal'' growth profile that is maximal between the crypt base and the crypt edges. This concept is illustrated in Fig.\ \ref{fig:growthstructurecryptdiagram}. 

The second aspect of interest concerns homeostasis. Generically, homeostasis refers to a target state of a system, an equilibrium that is usually thought of as optimal in some way for the functioning of the system. In growing tissues, homeostasis is characterised by a balance between cell division and cell death or extrusion \cite{Anonymous:xgTthBz1}, such that the morphological properties of the tissue (shape, size) do not change with time.  Homeostasis in the crypt is particularly interesting in this regard: the bimodal growth profile noted above is maintained during homeostasis, as is the deeply-invaginated crypt morphology. This dynamic homeostasis requires a delicate balance of growth, cell migration, and the extrusion or ``sloughing'' of cells at the top of the crypt. 

Numerous factors can contribute to growth regulation and the maintenance of a homeostatic state; these can be either chemical or physical. Our goal in this paper is to investigate the role of mechanics. Mechanical forces have been found to be a key contributor in growth regulation and in homeostasis in a number of systems \cite{legoff2016mechanical}. The principal idea in mechanically-driven growth and homeostasis is that growth of a tissue depends on the difference between the stress in the tissue and a target homeostatic stress. For instance, parts of the tissue that are in relative tension compared to the target stress will grow to relieve the tension. In purely-mechanical growth, mechanical homeostasis occurs when the stress is exactly equal to the target stress, at which point growth is halted \cite{taber2009towards, erlich2019homeostatic}. Mechanically-driven growth may also be combined with other cues, (for example, biochemical) so that mechanical forces enhance or reduce the growth rate \cite{ERLICH:2018di}. It is this latter case that is of interest here, with Wnt signalling acting as a well-known regulator of crypt proliferation \cite{spit2018tales}.

In particular, we consider two questions: 
\begin{enumerate}
\item[(1)] Can growth driven by a unimodal biochemical signal (Wnt), but regulated by mechanical feedback, produce a ``bimodal'' proliferative structure?
\item[(2)] What are the conditions on the system for dynamic homeostasis to be maintained, and can this be achieved consistently with (1)?
\end{enumerate}

While these questions are motivated specifically by the crypt, similar questions may be relevant to a number of related systems. In a broader sense, the issues considered are: (i) whether mechanical feedback can qualitatively alter biochemical patterns of growth, and (ii) how to approach the mechanics of non-static tissue homeostasis. A key objective here is to formulate a modelling framework capable of treating such issues. While patterns generated solely through biochemical processes, otherwise known as Turing patterns, have been well studied and continue to be an active area of research \cite{crampin1999reaction, krause2020one}, we focus on mechanical pattern formation in this work, where the interplay between growth and stress drive the transition from a trivial, flat morphology to a non-trivial, buckled morphology. 

Our approach will be to investigate these questions in a continuum setting. We model the cross-section of a single crypt, treating the line of epithelial cells as a growing, elastic rod. Similar models have appeared in the literature \cite{Edwards2007, Nelson2011}, in which the mechanics emerges from first principles and the resulting system is defined by partial differential equations, for which numerous computational and analytical tools are available. At the same time, considerable care must be taken when incorporating cellular-level processes such as sloughing and localised growth within a continuum framework. In terms of dynamic homeostasis, a subtle issue arises in even defining homeostasis, given that growth is not halted and, at the level of cells, there is a ``treadmilling process''. Our starting point is to define homeostasis from a biologist's perspective: an observer watching any particular point along the crypt would see a fixed rate of cell division and cell migration at all times, without any change in morphology. As we show, translating this concept to a precise definition in a continuum mechanics setting requires careful delineation of variables. While solid mechanics is most naturally expressed using Lagrangian variables, we show that a characterisation of dynamic homeostasis requires translation of the governing system to an Eulerian representation. In this way, we derive the necessary conditions relating growth rate, migration velocity, and sloughing rate for homeostasis to exist.

This paper is structured as follows. In Section \ref{sec:modellingframework}, we present the mathematical framework for modelling dynamical tissue homeostasis that is of the type mentioned above. 
For a given form of mechano-chemical growth, we identify necessary conditions for homeostasis to occur in Section \ref{sec:homeostasisconditions}. In Section \ref{sec:formofmechanicalfeedback}, we identify functional forms of mechanical feedback that generate the proliferative structure of the crypt, by considering the system in the absence of curvature. Returning to the original, 2D morphology, we compute the homeostatic states and analyse their dynamic stability in Section \ref{sec:computinghomeostasis}. We end this paper with a discussion of results and possible extensions in Section \ref{sec:discussion}.

\section{Modelling framework}
\label{sec:modellingframework}
We first outline the mathematical framework used to model the crypt. The colonic crypt comprises a highly-deformed morphology, in which various biochemical and biomechanical factors contribute to its morphogenesis and homeostasis. A biologically-realistic description of the crypt must capture the interplay between local tissue growth, which we assume to be driven by both chemical and mechanical cues. The former is modelled as an ever-present Wnt signal profile. The latter is strongly linked to the local environment, comprising the basement membrane, to which the crypt is anchored, and the surrounding non-epithelial tissue stroma \cite{meran2017intestinal}. We exploit several modelling assumptions.

As the crypt shape is similar to that of a test tube, there is an approximate radial symmetry about the crypt base. Therefore, we can consider the crypt geometry from a cross-sectional view, as if one has taken a histological slice of the tissue, and we can model the transverse deformations. This allows for a convenient 1D parametrisation of the crypt epithelium, which we model as a growing line of cells deforming within the $x-y$ plane. As the length of the crypt epithelium is much greater than the height and width of a single cell, this slenderness ratio allows us to adopt a continuum approach, representing the proliferating line of cells as a growing, elastic rod embedded in a plane. 

Rather than model both the supporting basement membrane and tissue stroma explicitly, we abstract the mechanical effects of this composite material into a single force density applied along the epithelial line. We follow the approach of a foundation force, often employed in elastic rod models \cite{Edwards2007, Nelson2011, Moulton2013, Chirat2013}, representing the attachment of the rod to an underlying substrate, but adapt the foundation model here to allow remodelling of the stroma and basement membrane.

\begin{figure*}[t!]
	\centering
	\includegraphics[width=0.8\textwidth]{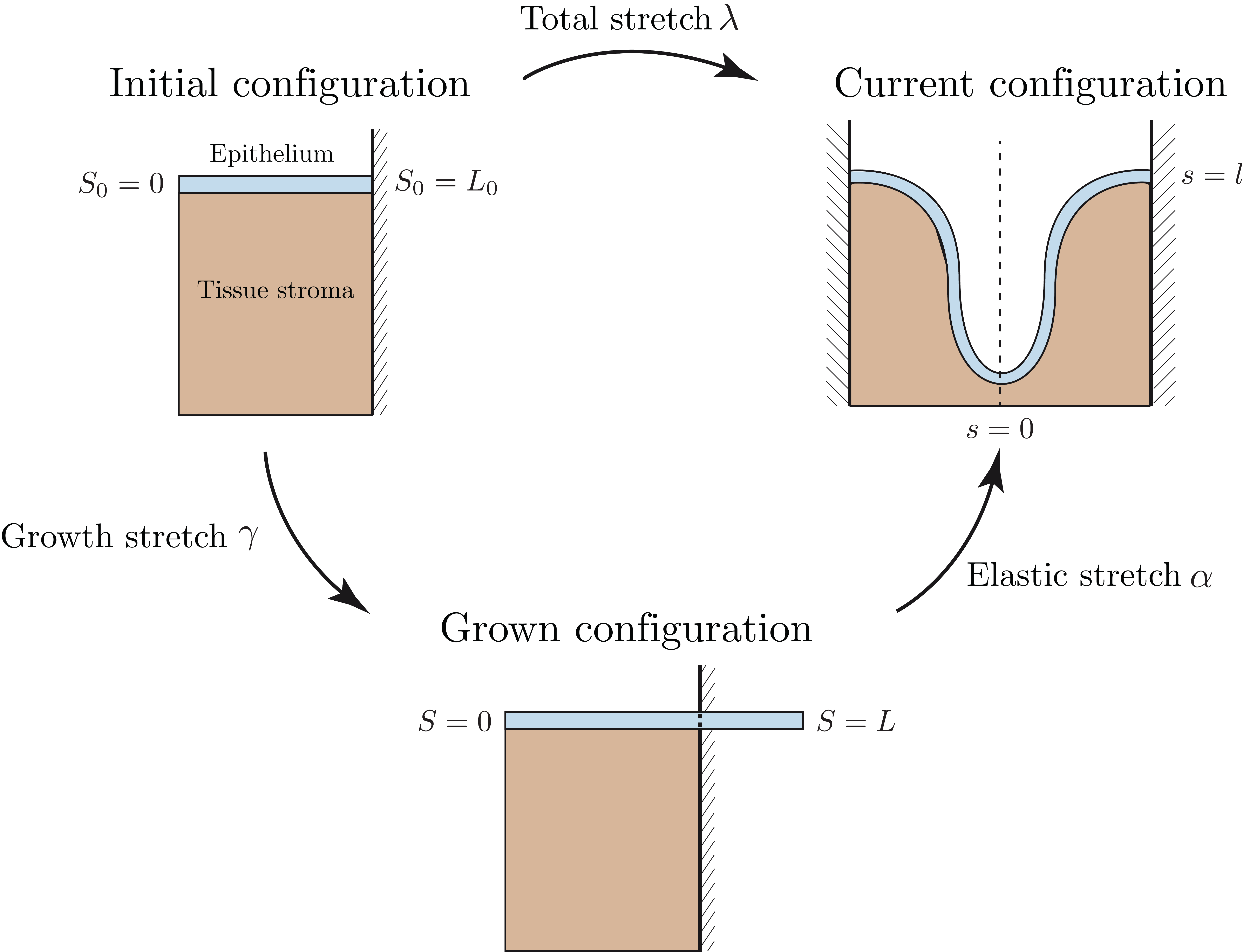}
	\caption{\textbf{The morphoelastic rod approach, applied to the crypt.} The blue region indicates the crypt epithelium, modelled as a growing, elastic rod, while the brown region represents the non-epithelial tissue stroma. Note that by crypt shape symmetry, we need only deformations along the half-interval, from the crypt base to the top. }
	\label{fig:morphoroddiagram}
\end{figure*}

In order to focus on mechanical effects, we take a simplistic approach to the biochemistry, assuming that there is a prescribed background concentration of Wnt present at each point along the rod. The growth of the rod is taken to be due to a combination of Wnt concentration and mechanical stress. We also assume that growth occurs on a much slower timescale than that of any elastic deformations, so that the system is always in quasi-static mechanical equilibrium. For this setup, we work within the framework of morphoelastic rods \cite{Moulton2013}, in which the growth and elastic deformation of the rod are defined by three distinct configurations: the initial, pre-grown Lagrangian configuration, which is stress-free and parametrised by the initial arc length, $S_0$; the grown configuration, parametrised by the grown arc length, $S$, which is a virtual configuration and still stress-free; and the current Eulerian configuration, parametrised by the current arc length, $s$. 
By the morphoelasticity assumption, the rod arc length evol-ves due to a growth process, characterised by the growth stretch, $\gamma(S_0, t) = \partial S/\partial S_0$, and subsequently due to an elastic deformation, modelled by the elastic stretch, $\alpha = \partial s/\partial S$. Therefore, the total stretch, $\lambda = \partial s/\partial S_0$, from the initial to the current configuration can be expressed as: 
\begin{align}
\lambda = \alpha\gamma \qquad \Longleftrightarrow \qquad  \frac{\partial s}{\partial S_0} = \frac{\partial s}{\partial S}\frac{\partial S}{\partial S_0}.\label{eq:morphoelasticityassumption}
\end{align}
This principle is illustrated in Fig.\ \ref{fig:morphoroddiagram}.

\subsection{Geometry and mechanics}
We now outline the governing equations for the growing rod. As growth is most naturally defined as a function of the initial arc length $S_0$, we first describe the governing equations with respect to $S_0$. 

The rod shape is described by its centreline, modelled as a 2D curve $\mathbf{r}(S_0, t) = x\mathbf{e}_x + y\mathbf{e}_y$. Let $\theta$ denote the angle between the tangent vector $\boldsymbol{\tau}=\cos\theta\mathbf{e}_x + \sin\theta\mathbf{e}_y$ and the $x$-axis. Geometry supplies
\begin{align}
&\frac{\partial x}{\partial S_0} = \alpha\gamma\cos\theta,\label{eq:dim2Dx}\\
&\frac{\partial y}{\partial S_0} = \alpha\gamma\sin\theta.\label{eq:dim2Dy}
\end{align}
Note that the factor $\alpha\gamma$ arises when parametrising with respect to $S_0$. 

 Let $\mathbf{n}(S_0, t) = n_x\mathbf{e}_x + n_y\mathbf{e}_y$ denote the resultant force within the rod. The mechanical effects of the supporting basement membrane and tissue stroma are modelled through a foundation force proportional to $(\mathbf{r}-\mathbf{p})$, where the 2D curve $\mathbf{p} = p_x\mathbf{e}_x + p_y\mathbf{e}_y$ denotes the position of the foundation. Unlike the classical Winkler foundation, here we allow the foundation position to evolve over time (i.e. $\mathbf{p}=\mathbf{p}(S_0,t)$), such that the foundation relaxes to the current rod shape at rate $\eta^{-1}$, similar to Chirat et al.\ \cite{Chirat2013}. This feature captures the remodelling of the underlying stroma and basement membrane in response to the deformation of the epithelium, and enables us to more realistically simulate large deformations than with a static foundation force \cite{Almet2019b}. In component form, the balance of linear momentum and foundation relaxation read:
\begin{align}
&\frac{\partial n_x}{\partial S_0} = \alpha\gamma Ek_f(x - p_x), \qquad \frac{\partial p_x}{\partial t}= \frac{1}{\eta}(x - p_x),\label{eq:dim2Dforcex}\\
&\frac{\partial n_y}{\partial S_0} = \alpha\gamma Ek_f(y - p_y), \qquad \frac{\partial p_y}{\partial t} = \frac{1}{\eta}(y - p_y).\label{eq:dim2Dforcey}
\end{align}
Here, $E$ is the Young's modulus of the rod, so that the dimensionless parameter $k_f$ relates foundation stiffness to rod stiffness. The factor $\alpha\gamma$ again appears because we express the system in Lagrangian form, so that the force density $Ek_f(\mathbf{r}-\mathbf{p})$ is  a force per current length.

Letting $\mathbf{m} = m\mathbf{e}_z$ denote the resultant rod moment, the balance of angular momentum is given by:
\begin{align}
&\frac{\partial m}{\partial S_0} = \alpha\gamma(n_x\sin\theta - n_y\cos\theta).\label{eq:dim2Dmoment}
\end{align}
The force and moment balance is supplemented by constitutive laws for bending and stretching. We relate the bending moment, $m$, to the flexure, $\partial\theta/\partial S$, through the standard relation:
\begin{align}
m = \frac{EI}{\gamma}\frac{\partial \theta}{\partial S_0},\label{eq:dimlagrangianbending}
\end{align}
where $I$ is the moment of inertia. Note that the moment is proportional to the flexure, $\partial\theta/\partial S$, rather than the curvature $\partial\theta/\partial s$, as $\partial\theta/\partial S$ is independent of any stretching within the rod. We also relate the elastic stretch $\alpha$ to the axial stress through a linear constitutive relation:
\begin{align}
&n_\tau:=\mathbf{n}\cdot\boldsymbol{\tau}=n_x\cos\theta + n_y\sin\theta = EA(\alpha - 1),\label{eq:dimelasticstretch}
\end{align}
equivalent to a Hookean spring, where $A$ is the area of the rod cross-section. The constitutive law \eqref{eq:dimelasticstretch} models the extensibility of the rod. Note that for an inextensible rod, \eqref{eq:dimelasticstretch} would be replaced by the geometric constraint $\alpha \equiv 1$.

Finally, the system is driven by imposing a growth law of the form:
\begin{align}
&\frac{\partial\gamma}{\partial t}=\gamma \,G(\text{Wnt},n_\tau, t, \dots),\label{eq:growthlawgeneric}
\end{align}
where the function $G$ could incorporate numerous effects, but in our analysis will depend only on Wnt concentration and axial stress $n_\tau$. 

It remains to impose boundary and initial conditions. A typical crypt morphology is symmetric about the base. Supposing that the full crypt encompasses the region $-L_0\leq x\leq L_0$, here we exploit this symmetry and consider a half domain, valid for rod morphologies symmetric about $x = 0$. Thus, we restrict attention to the domain $x\in [0, L_0]$, shifting the point $S_0=0$ to the middle of the rod and imposing a symmetry condition at $S_0=0$ and a clamped boundary condition at $S_0=L_0$:
\begin{align}
&x(0) = 0, \quad n_y(0) = 0, \quad \theta(0) = 0,\nonumber\\
&x(L_0) = L_0, \quad y(L_0) = 0, \quad \theta(L_0) = 0,\label{eq:dimBCs}
\end{align}
Natural initial conditions are a flat foundation and rod shape: $x(S_0,0)=p_x(S_0, 0) = S_0$ and $y(S_0, 0) =p_y(S_0, 0) = 0$, plus zero initial growth, $\gamma(S_0,0)=1$, in which case the force and moment are initially zero. While these conditions are needed in the context of morphogenesis, in our analysis of homeostasis we shall see that prescribing such conditions is not actually necessary.

\subsection{Non-dimensionalisation}	
To reduce the number of model parameters in the system, we non-dimensionalise independent and dependent variables in the following manner:

\begin{align}
t^* &= T t,\nonumber\\
(S_0^*, x^*, y^*, p_x^*, p_y^*) &= L_0(S_0, x, y, p_x, p_y),\nonumber\\
(n_x^*, n_y^*) &= EIL_0^{-2}(n_x, n_y),\nonumber\\
m^* &= EIL_0^{-1}m,\label{eq:nondimensionalisation}
\end{align}
where $T$ is the typical growth timescale. Substituting \eqref{eq:nondimensionalisation} into Equations \eqref{eq:dim2Dx}--\eqref{eq:growthlawgeneric} and dropping asterisks for notational convenience then yields the full nondimensional system:
\begin{align}
\frac{\partial x}{\partial S_0} &= \alpha\gamma\cos\theta,\label{eq:lagrangian2Dx}\\
\frac{\partial y}{\partial S_0} &= \alpha\gamma\sin\theta,\label{eq:lagrangian2Dy}\\
\frac{\dot\gamma}{\gamma} &=g= TG,\\
\frac{\partial n_x}{\partial S_0} &=\alpha\gamma k(x - p_x), \qquad \dot{p}_x= \rho(x - p_x),\label{eq:lagrangian2Dforcex}\\
\frac{\partial n_y}{\partial S_0} &= \alpha\gamma k(y - p_y), \qquad \dot{p}_y = \rho(y - p_y),\label{eq:lagrangian2Dforcey}\\
\frac{\partial \theta}{\partial S_0} &= \gamma m,\label{eq:lagrangian2Dtheta}\\
\frac{\partial m}{\partial S_0} &= \alpha\gamma(n_x\sin\theta - n_y\cos\theta),\label{eq:lagrangian2Dmoment}\\
n_\tau& = \mathcal{S}^{-1}(\alpha - 1),\label{eq:lagrangianelasticstretch}
\end{align}
where $k$ is the (non-dimensional) foundation stiffness and $\rho$ is the ratio of the growth timescale to the remodelling timescale of the foundation attachments. A larger value of $k$ indicates a stiffer foundation, while larger values of $\rho$ correspond to more rapid relaxation of the foundation. Also, $\mathcal{S}$ is the ``stretchability'' of the rod, measuring the ratio of the bending stiffness to the stretching stiffness; the case $\mathcal{S} = 0$ corresponds to an inextensible rod \cite{pandey2014dynamics}. 

The boundary conditions \eqref{eq:dimBCs} rescale to:
\begin{align}
&x(0) = 0, \quad n_y(0) = 0, \quad \theta(0) = 0,\nonumber\\
&x(1) = 1, \quad y(1) = 0, \quad \theta(1) = 0.\label{eq:lagrangianBCs}
\end{align}
The dimensionless equations \eqref{eq:lagrangian2Dx}--\eqref{eq:lagrangianelasticstretch} contain three model parameters, $k$, $\rho$, and $\mathcal{S}$. Assuming a rectangular cross-section with height $h$ and width $w$, these parameters are given by:
\begin{align}
k = \frac{12k_fL_0^4}{wh^3}, \qquad \rho = \frac{T}{\eta}, \qquad \mathcal{S} = \frac{h^2}{12L_0^2}.\label{eq:nondimparameters}
\end{align}
Based on estimates from human colonic crypt histologies \cite{taylor2003mitochondrial}, we fix $w = 10\mu$m, $h = 15\mu$m, and $L_0 = 125\mu$m. We also set the growth timescale $T = 24$ hours to reflect the timescale of tissue morphogenesis. Later, we will $k_f = 0.01$ such that the crypt shape is even about $x = 0$ and contains a single invagination and fix $\rho = 10$, corresponding to rapid relaxation of the foundation to the rod shape, increasing the invagination depth \cite{Almet2019b}.
\subsection{Homeostasis definition and framework}

Before proceeding, it is important to define what is meant by homeostasis. As described in the Introduction, this will depend on the form of growth law, given by the function, $g$. In the case of purely mechanical growth towards a homeostatic axial stress, $n_\tau^*$, we would have $g=f(n_\tau-n_\tau^*)$, where $f$ is a function with the properties $f(0)=0$, $f(x)>0$ for $x>0$, and $f(x)<0$ for $x<0$, which models enhanced growth due to relative tension and the growth inhibition due to relative compression. In this case homeostasis corresponds to the state with $n_\tau=n_\tau^*$ for all $S_0$ and thus $\dot\gamma=0$ (as well as a static foundation). Such a state would truly be static, with all variables unchanging in time. More interesting, and relevant to the crypt, is a growth law that combines biochemical and biomechanical signals. Letting $W=W(S_0,t)$ denote the concentration of Wnt, we consider the generic additive form 
\begin{equation}
g=\mu_1W+f(n_\tau-n_\tau^*),
\end{equation}
where $\mu_1$ denotes the sensitivity of proliferation to Wnt. Since we are decoupling the biochemistry we can incorporate $\mu_1$ into $W$ and thus set $\mu_1=1$. 

Our baseline assumption is that the Wnt signal persists throughout time, and has a fixed functional form {\it in the Eulerian frame}. That is, Wnt can be viewed as a property of distance from the crypt base, described by the Eulerian variable $s$, as opposed to being a material property, described by the Lagrangian variable $S_0$. This reflects the notion that any biochemical process---for example, the diffusion of morphogens---is occurring in current position of the tissue. Therefore, we prescribe a form $W=W(s)$.

\begin{figure*}
\centering
\includegraphics[width=\textwidth]{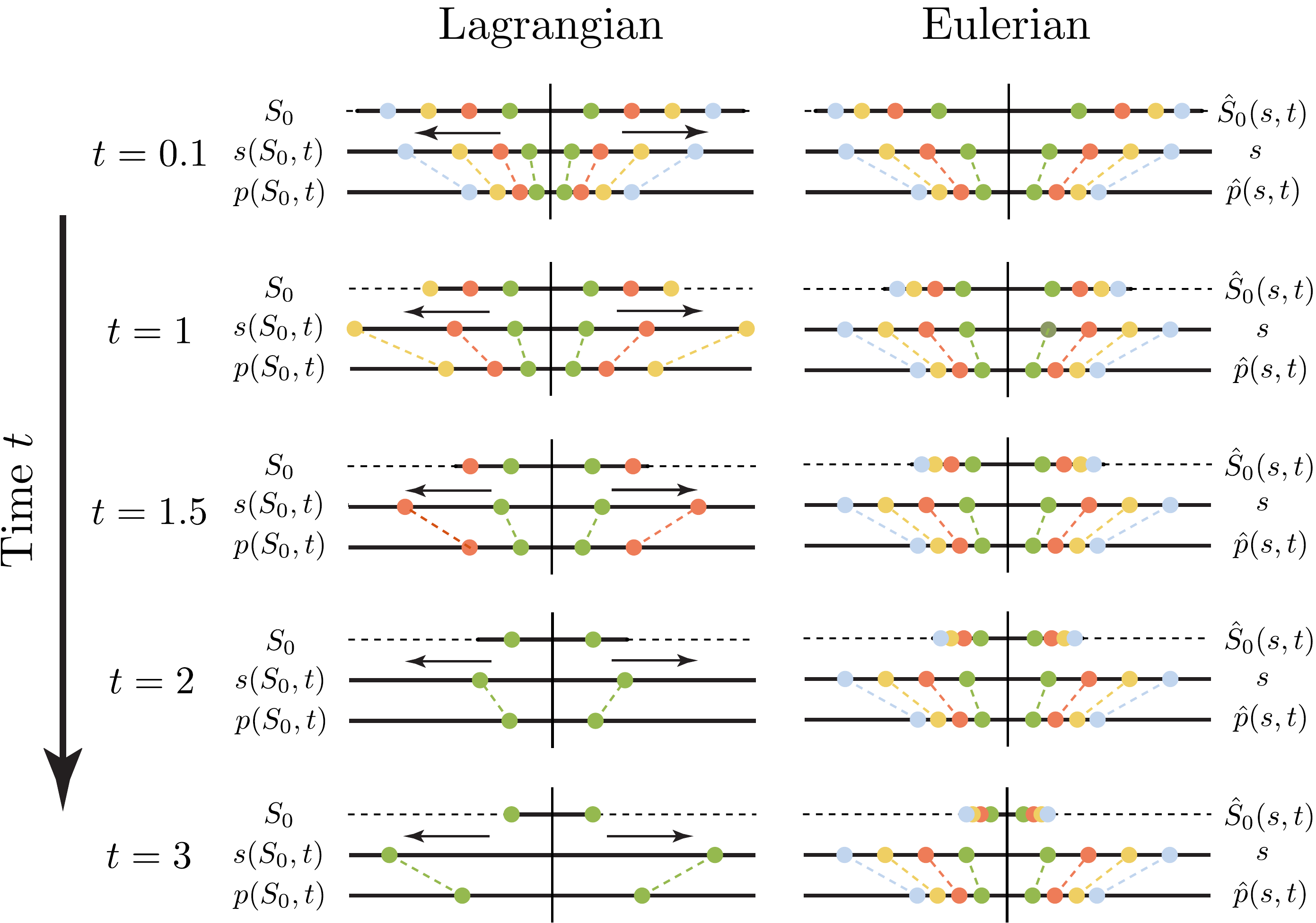}
\caption{\textbf{Demonstration of homeostasis in Lagrangian and Eulerian frames}. The initial and current arc lengths, as well as the foundation positions, are plotted for growth on a 1D line, simulating homeostasis in the crypt. On a 1D geometry, the foundation attachment positions are simply given by $p(S_0, t) := p_x(S_0, t)$. Left column: the Lagrangian representation, with initial material arc length $S_0$ the independent variable. The marked circles correspond to $S_0 = 0.2$ (green), $S_0 = 0.4$ (orange), $S_0 = 0.6$ (yellow), and $S_0 = 0.8$ (blue) and their even extensions. Right column: the Eulerian representation, with current arc length $s$ the independent variable, and circles at fixed $s = 0.2, 0.4, 0.6, 0.8$ (green, orange, yellow and blue circles respectively) and even extensions. The evolution of the sloughing boundary $S_0 = L_\mu(t)$ is denoted by the dashed lines. In the Lagrangian configuration, material points move outward until they are sloughed away and disappear; in the Eulerian configuration, the profiles of $s$ and $\hat p$ are fixed, and the map back to $S_0$ shows points converging to the centre as the $S_0$ domain shrinks.}
\label{fig:eulerianvslagrangianhomeostasis}
\end{figure*}

Given this, for growth to halt, the mechanosensitive term would need to cancel the Wnt signal exactly. In practice, however, homeostasis in the crypt is observed to be dynamic, a non-homogeneous treadmilling state, with growth persisting and balanced by sloughing of material at the top of the crypt. What {\it is} static in homeostasis is the growth rate and material velocity at any point $s$, as well as the morphology. In terms of experimental observations, a biologist watching a fixed point in space (in the Eulerian frame) will observe a constant rate of cell division and cell migration, and an unchanging crypt shape. Such a static state of growth and velocity in the Eulerian frame is {\it not static} in the Lagrangian frame. This concept is illustrated for a 1D geometry in Fig.\ \ref{fig:eulerianvslagrangianhomeostasis}, where we compare (in a simplified 1D morphology for illustration) a homeostatic process viewed in the Lagrangian and Eulerian frames. In the Eulerian view, with current arc length $s$ as the independent variable, the current position and foundation attachments do not change with time, while the pre-image of each point, i.e the material point $\hat{S}_0(s,t)$, continually moves inward. By contrast, in the Lagrangian view, with material point $S_0$ as the independent variable, there is a clear migration outward of the current points $s(S_0, t)$, as well as the corresponding foundation attachments, to the edge of the domain where they are ultimately removed from (sloughed out of) the Lagrangian domain.  Thus, in homeostasis, while variables are fixed with respect to the Eulerian configuration, the Lagrangian configuration evolves continuously to maintain the homeostatic growth profile. 

To express this notion of homeostasis mathematically, we first note that in the morphoelastic rods framework, growth rate is described by the incremental growth\footnote{The growth stretch of a small section over a time increment $\Delta t$ is given by $1+\dot\gamma/\gamma\Delta t$. Contrast this with the function $\gamma$ itself, which describes the change in arc length at a given material point over the entirety of the growth history.} $\dot\gamma\gamma^{-1}$, i.e. the function $g$, while the material velocity is $v=\partial{s}/\partial{t}$ and the morphology is fully determined by the function $\theta$. Homeostasis can then be defined by the condition that $g$, $v$ and $\theta$ are functions of $s$ only (and not time $t$). To characterise homeostasis, it is thus prudent to cast the system in an Eulerian frame. In order to define quantities correctly, we must take particular care with time derivatives. To clarify the notation, we will use hats for a quantity expressed in Eulerian form. That is, for a function expressed in Lagrangian form $f(S_0, t)$, we denote its equivalent Eulerian form by $\hat{f}(s, t) := f(S_0(s, t), t)$. By the chain rule and the multiplicative decomposition \eqref{eq:morphoelasticityassumption}, the associated space and time partial derivatives are:
\begin{align}
&\frac{\partial\hat{f}}{\partial s} = \frac{\partial f}{\partial S_0}\frac{1}{\alpha\gamma},\label{eq:materialspacederivative}\\
&\frac{\partial\hat{f}}{\partial t} = \frac{\partial f}{\partial t} + \hat{v}\frac{\partial f}{\partial s},\label{eq:materialderivative}
\end{align}
where the (pull-back) velocity $\hat{v}(s, t)$ is given by
\begin{align}
\hat{v}(s, t) = \hat{\alpha}\hat{\gamma}\frac{\partial \hat{S}_0}{\partial t}.\label{eq:flowvelocity}
\end{align} 

%

The material velocity $v(S_0, t) = \partial s(S_0, t)/\partial t$, measures the velocity at a fixed material point $S_0$---the continuum form of a cell migration velocity. The migration velocity is routinely measured in the crypt and provides a typical quantitative measure of homeostasis \cite{kaur1986cell, krndija2019active}. Applying the chain rule to $f(S_0, t) = \hat{f}(s(S_0,t), t)$ gives
\begin{align}
\frac{\partial f}{\partial t} = \frac{\partial\hat{f}}{\partial t} + v\frac{\partial\hat{f}}{\partial s}.\label{eq:lagrangiantimederivative}
\end{align}
Rearranging \eqref{eq:lagrangiantimederivative} for $\partial\hat{f}/\partial t$ and equating this definition with Equation \eqref{eq:materialderivative} implies that the two velocities are related as follows:
\begin{align}
v(S_0, t) = -\hat{v}(s(S_0, t), t).\label{eq:currentvelocitydefn}
\end{align}

In the Eulerian frame, the spatial domain is $s \in [0, l]$, where $l=l(t)$ is the current rod length, satisfying
\begin{align}
l= \int^{1}_0\alpha\gamma dS_0.\label{eq:totalrodlength}
\end{align}
Following the conversions above, growth in the Eulerian configuration evolves according to the partial differential equation,
\begin{align}
\frac{\partial\hat{\gamma}}{\partial t} = g(s,t)\hat{\gamma} + \hat{v}\frac{\partial\hat{\gamma}}{\partial s}.\label{eq:euleriangrowth}
\end{align}
From the multiplicative decomposition \eqref{eq:morphoelasticityassumption}, we can compute the Eulerian form of the initial arc length $\hat{S}_0(s, t)$ via
\begin{align}
&\frac{\partial \hat{S}_0}{\partial s} = \frac{1}{\hat{\alpha}\hat{\gamma}}.\label{eq:eulerianmorphoelasticityassumption}
\end{align}
Applying \eqref{eq:materialspacederivative}--\eqref{eq:materialderivative} to the Lagrangian system \eqref{eq:lagrangian2Dx}-\eqref{eq:lagrangianelasticstretch}, the remaining equations read:
\begin{align}
\frac{\partial \hat{x}}{\partial s} &= \cos\hat{\theta},\label{eq:eulerian2Dx}\\
\frac{\partial\hat{y}}{\partial s} &= \sin\hat{\theta},\label{eq:eulerian2Dy}\\
\frac{\partial \hat{n}_x}{\partial s} &= k(\hat{x} - \hat{p}_x), \quad \frac{\partial\hat{p}_x}{\partial t} = \rho(\hat{x} - \hat{p}_x) + \hat{v}\frac{\partial \hat{p}_x}{\partial s},\label{eq:eulerian2Dforcex}\\
\frac{\partial \hat{n}_y}{\partial s} &= k(\hat{y} - \hat{p}_y), \quad \frac{\partial \hat{p}_y}{\partial t} = \rho(\hat{y} - \hat{p}_y) + \hat{v}\frac{\partial \hat{p}_y}{\partial s},\label{eq:eulerian2Dforcey}\\
\frac{\partial \hat{\theta}}{\partial s} &= \frac{\hat{m}}{\hat{\alpha}},\label{eq:eulerian2Dtheta}\\
\frac{\partial \hat{m}}{\partial s} &= \hat{n}_x\sin\hat{\theta} - \hat{n}_y\cos\hat{\theta},\label{eq:eulerian2Dmoment}\\
\hat{n}_\tau &= \mathcal{S}^{-1}\left(\hat{\alpha} - 1\right),\label{eq:eulerianelasticstretch}
\end{align}
while the Eulerian boundary conditions take the form
\begin{align}
&\hat{x}(0) = 0, \quad \hat{n}_y(0) = 0, \quad \hat{\theta}(0) = 0,\nonumber\\
&\hat{x}(l) = 1, \quad \hat{y}(l) = 0, \quad \hat{\theta}(l) = 0.\label{eq:eulerianBCs}
\end{align}

The Eulerian formulation requires care with two issues. One is that additional spatial derivatives have appeared, which must be balanced by extra boundary conditions. These relate to the velocity at the left boundary, $s = 0$, and are outlined below. The other issue concerns $\hat{S}_0(s, t)$. Observe that there are two first-order partial differential equations for $\hat{S}_0$, \eqref{eq:flowvelocity} and \eqref{eq:eulerianmorphoelasticityassumption}. To ensure that $\hat{S}_0$---and, consequently, the velocity $\hat{v}(s, t)$---are defined consistently, we require the following compatibility condition:
\begin{align}
\frac{\partial^2\hat{S}_0}{\partial s\partial t} = \frac{\partial^2\hat{S}_0}{\partial t \partial s} \ \Longleftrightarrow\ \frac{\partial}{\partial s}\left(\frac{\hat{v}}{\hat{\alpha}\hat{\gamma}}\right) = \frac{\partial}{\partial t}\left(\frac{1}{\hat{\alpha}\hat{\gamma}}\right),\label{eq:S0compatibility}
\end{align}
which connects the velocity to the growth and stretch. 

The Eulerian system as outlined above, including the compatibility equation, \eqref{eq:S0compatibility}, holds in general, regardless of whether the system is in homeostasis or not. For most applications, it would be disadvantageous to solve the system in Eulerian form due to the changing spatial domain and extra derivatives and boundary conditions. For a dynamic homeostasis, however, the Eulerian formulation enables us to identify the conditions necessary for a homeostatic state. 

Before addressing the specifics of homeostasis, the final step needed in our framework is a description of sloughing, i.e. the loss or extrusion of cells at the top of the crypt. If growth persists without changing the morphology, material must be lost. A loss of material at the boundary $s=l$ is equivalent to reducing the Lagrangian domain. 
To this end, we define the sloughing boundary
\begin{align}
L_\mu(t) := \hat{S}_0(l, t) =  \int^l_0\frac{1}{\hat{\alpha}\hat{\gamma}}ds.\label{eq:sloughingboundary}
\end{align}
At any time, the material domain is $\hat{S}_0\in[0,L_\mu(t)]$, so that the region $(L_\mu,1]$ is effectively removed from the system.
If, for instance, continued growth occurs (i.e. $\hat\gamma(s, t)$ is an increasing function of time) in an inextensible rod ($\hat\alpha\equiv1$), and the current length $l$ does not vary in time, \eqref{eq:sloughingboundary} shows that the Lagrangian domain will shrink monotonically. With the definition of $L_\mu(t)$ from \eqref{eq:sloughingboundary}, which in itself is not particularly biologically meaningful, we will show in Sect.\ \ref{sec:growthsloughinghomeostasis} how one can derive a more relevant sloughing \emph{rate} that measures the net cell turnover rate.

\section{Conditions for homeostasis}
\label{sec:homeostasisconditions}
We now proceed to derive the necessary conditions for homeostasis. The starting point is the definition that incremental growth, morphology, and velocity, in Eulerian form, are all functions of $s$ only. Thus, considering the mechanochemical incremental growth:
\begin{align}
g(s,t) = W(s) + f(\hat{n}_\tau - \hat{n}^*),\label{eq:homeostasisincgrowth}
\end{align}
if $g=g(s)$ in homeostasis, it follows that $\hat n_\tau=\hat n_\tau(s)$. Since the shape is completely determined by $\hat\theta=\hat\theta(s)$ in homeostasis, then by the definition $\hat n_\tau=\hat n_x\cos\hat\theta+\hat n_y\sin\hat\theta$, it must also be true that $\hat{n}_x$ and $\hat{n}_y$ are independent of time:
\begin{align}
\hat{n}_\tau &= \hat{n}_\tau(s), \,\hat{\theta} = \hat{\theta}(s),\nonumber\\
&\Longrightarrow\ \hat{n}_x = \hat{n}_x(s), \,\hat{n}_y = \hat{n}_y(s).\label{eq:homeostasiselasticstretch}
\end{align}
By the constitutive relation \eqref{eq:eulerianelasticstretch}, the Eulerian elastic stretch $\hat{\alpha}$ is also independent of time:
\begin{align}
\hat{\alpha}(s) = 1 + \mathcal{S}\hat{n}_\tau(s).
\end{align} 
Next, note that $(\hat{x}, \hat{y})$ can be obtained straightforwardly from $\hat\theta(s)$:
\begin{align}
&\hat x=\hat{x}(s) = \int^s_0\cos\hat{\theta}(\xi)d\xi,
\\&\hat y=\hat{y}(s) = \int^s_0\sin\hat{\theta}(\xi)d\xi - \int^l_0\sin\hat{\theta}(\xi)d\xi.\label{eq:homeostasis2Dshape}
\end{align}

Equations \eqref{eq:eulerian2Dforcex}--\eqref{eq:eulerian2Dforcey} then imply that $\hat{p}_x = \hat{p}_x(s)$ and $\hat{p}_y = \hat{p}_y(s)$. Furthermore, by Equations \eqref{eq:eulerian2Dmoment} and \eqref{eq:homeostasiselasticstretch}, the bending moment $\hat{m} = \hat{m}(s)$.

Setting the time derivatives, $\partial\hat{p}_x/\partial t$ and $\partial\hat{p}_y/\partial t$, to zero, we can then simplify the foundation relaxation equations to yield the following pair of first-order (in space) ordinary differential equations (ODEs):
\begin{align}
&\hat{p}_x' = \frac{\rho(\hat{p}_x - \hat{x})}{\hat{v}},\label{eq:homeostasiseqnspx}\\
&\hat{p}_y' = \frac{\rho(\hat{p}_y - \hat{y})}{\hat{v}},\label{eq:homeostasiseqnspy}
\end{align}
where the prime $'$ denotes differentiation with respect to $s$. We can also obtain a first-order ODE for $\hat v(s)$ from the compatibility condition \eqref{eq:S0compatibility}. Using $\partial\hat{\alpha}/\partial t = 0$, Equation \eqref{eq:S0compatibility} simplifies to
\begin{align}
\hat{v}' = \frac{\hat{\alpha}'}{\hat\alpha}\hat{v} - g(s;\hat n_\tau)\label{eq:velocityinhomeostasis}.
\end{align}
Here, we write $g=g(s;\hat n_\tau)$ to make explicit that the axial stress profile is incorporated in the functional form of $g$.

To summarise, in homeostasis, the variables
\begin{align}
\mathbf{U}=\{\hat x,\hat y,\hat\theta,\hat n_x,\hat n_y, \hat\alpha, \hat p_x,\hat p_y,\hat m,\hat v\}\label{eq:timeindependentsolns}
\end{align}
are all functions of $s$ only. Note this set does not include the growth $\hat\gamma$ nor the initial arc length $\hat S_0$, but this is to be expected: in dynamic homeostasis, growth persists and thus depends on time; it follows that the map from current to initial arc length does as well. Moreover, the set of first-order ODEs for $\mathbf{U}(s)$ decouple from $\hat\gamma$ and $\hat S_0$, which evolve in homeostasis according to:
\begin{align}
\dot{\hat{\gamma}} &= g(s; \hat{n}_\tau(s))\hat{\gamma} + \hat{v}(s)\hat{\gamma}',\label{eq:growthhomeostasis}\\
{\hat S_0}'&=\frac{1}{\hat\alpha(s)\hat\gamma(s,t)}\label{eq:S0homeostasis}.
\end{align}

How to interpret the homeostatic system? The ODEs for $\mathbf{U}$ govern the relation between the incremental growth and velocity profiles ($g(s)$ and $\hat v(s)$) and the Eulerian profiles for shape, force, moment, stretch, and foundation position. Equations \eqref{eq:growthhomeostasis}-\eqref{eq:S0homeostasis} dictate how the growth and arc length maps must evolve in time to maintain those profiles. 
We solve this system computationally in Sect.\ \ref{sec:computinghomeostasis}. 

Counting variables, there are ten spatial variables in $\mathbf{U}$, but as $\hat\alpha$ is known constitutively, so there are only nine derivatives. At $s=0$, there are conditions on $\hat x$, $\hat\theta$, and $\hat n_y$, and conditions on $\hat x$, $\hat\theta$, and $\hat y$ at $s=l$,  providing only six conditions. However, in the Eulerian frame there are ``additional'' spatial derivatives on $\hat v$, $\hat p_x$, and $\hat p_y$. By symmetry, the velocity at the centre must vanish, so $\hat{v}(0) = 0$. Consequently, from Equations \eqref{eq:homeostasiseqnspx}--\eqref{eq:homeostasiseqnspy}, we must have that $\hat{p}_x(0) = \hat{x}(0) = 0$ and $\hat{p}_y(0) = \hat{y}(0)$, respectively, so that $
\hat{p}_x'(s)$ and $\hat{p}_y'(s)$ are not singular at $s = 0$, thus giving an additional three conditions. 

Therefore, in homeostasis, to obtain the spatial variables, $\mathbf{U}$, we solve Equations \eqref{eq:eulerian2Dx}--\eqref{eq:eulerian2Dmoment}, and Equations \eqref{eq:homeostasiseqnspx}--\eqref{eq:velocityinhomeostasis}, subject to the nine boundary conditions:
\begin{align}
&\hat{x}(0) = 0, \quad \hat{n}_y(0) = 0, \quad \hat{\theta}(0) = 0,\nonumber\\
&\hat{p}_x(0) = 0, \quad \hat{p}_y(0) = \hat{y}(0), \quad\hat{v}(0) = 0,\nonumber\\
&\hat{x}(l) = 0, \quad \hat{y}(l) = 0,\quad \hat\theta(l) = 0.\label{eq:homeostasisbcs}
\end{align}

\subsection{Growth and sloughing dynamics in homeostasis}
\label{sec:growthsloughinghomeostasis}
As shown, the time-independent variables $\mathbf{U}$ satisfy a boundary value problem (BVP) that can be solved independently from the variables $\hat\gamma$ and $\hat S_0$. Supposing such a solution has been found, we now construct a closed form solution for $\hat\gamma$ and $\hat S_0$, and use this to identify the rate of sloughing at the boundary needed for a consistent dynamic homeostasis.

From the definition of $\hat v$ in Equation \eqref{eq:flowvelocity}, we see that in homeostasis, $\hat v(s)\hat\alpha(s)^{-1}=\hat\gamma(s,t)\partial\hat{S}_0(s,t)/\partial t$. Since the left hand side is independent of $t$, the right hand side must be as well, implying that $\partial\hat{S}_0/\partial t$ and $\hat{\gamma}$ decompose into separable functions of $s$ and $t$ with cancelling time components. In particular, we have:
\begin{align}
\frac{\partial\hat{S}_0}{\partial t}= \frac{\Sigma(s)}{T(t)}, \qquad \hat{\gamma} = \Gamma(s)T(t). \label{eq:gammaansatz}
\end{align}
We first solve for $\hat{\gamma}$ by substituting the form \eqref{eq:gammaansatz} into Equation \eqref{eq:growthhomeostasis}. This yields the following:
\begin{align}
\frac{\dot{T}}{T} = g(s) + \hat{v}(s)\frac{\Gamma'}{\Gamma} = \beta \in \mathbb{R}.\label{eq:GammaT}
\end{align}
Solving for $T$ and $\Gamma$ respectively yields the solutions	
\begin{align}
&T(t) = T_0\mathrm{e}^{\beta t},\\
&\Gamma(s) = \Gamma_0\ \mathrm{exp}\left(\int^s_0\frac{\beta - g(s')}{\hat{v}(s')}ds'\right).
\end{align}
We note from the boundary condition $\hat{v}(0) = 0$ and \eqref{eq:GammaT} that  $\beta$ is given by
\begin{align}
\beta = g(0).\label{eq:betacondition}
\end{align}
Thus, we have an explicit form for growth in homeostasis:
\begin{align}
\hat{\gamma}(s, t) = \hat{\gamma}_0\ \mathrm{exp}\left(g(0)t + \int^s_0\frac{g(0) - g(s')}{\hat{v}(s')}ds'\right),\label{eq:gammasolhomeostasis}
\end{align}
where $\hat{\gamma}_0 = \Gamma_0T_0$ is a constant.

Now, the solution \eqref{eq:gammasolhomeostasis} implies that $\partial\hat{S}_0/\partial t$ takes the form
\begin{align}
\frac{\partial\hat{S}_0}{\partial t} = \Sigma(s)\mathrm{e}^{-g(0)t}.\label{eq:S0ansatz}
\end{align}
Substituting \eqref{eq:S0ansatz} and \eqref{eq:gammasolhomeostasis} into Equation \eqref{eq:flowvelocity} implies that 
\begin{align}
\Sigma(s) = -\frac{\hat{v}(s)}{g(0)\hat{\alpha}(s)\Gamma(s)}.\label{eq:Sigmasolhomeostasis}
\end{align}
We substitute the solution \eqref{eq:Sigmasolhomeostasis} into the form \eqref{eq:S0ansatz} and integrate with respect to time, obtaining the general solution for $\hat{S}_0$:
\begin{align}
\hat{S}_0(s, t) &=  -\frac{\hat{v}(s)\mathrm{exp}\left(-\int^s_0\frac{g(0) - g(s')}{\hat{v}(s')}ds'\right)}{g(0)\hat{\gamma}_0\hat{\alpha}(s)}\mathrm{e}^{-g(0)t}.\label{eq:S0solhomeostasis}
\end{align}
We remark that the constant that arises by integrating \eqref{eq:S0ansatz} vanishes due to the boundary condition $\hat{S}_0(0, t) = 0$. With our solutions to $\hat{\gamma}(s, t)$ and the homeostatic elastic stretch, $\hat{\alpha} = 1 + \mathcal{S}\hat{n}_\tau$, we can now calculate the sloughing boundary $L_\mu$ by evaluating \eqref{eq:S0solhomeostasis} at $s = l$:
\begin{align}
L_\mu(t) =-\frac{\hat{v}(l)\mathrm{exp}\left(-\int^l_0\frac{g(0) - g(s)}{\hat{v}(s)}ds\right)}{g(0)\hat{\gamma}_0(1 + \mathcal{S}\hat{n}_\tau(l))}\mathrm{e}^{-g(0)t}.\label{eq:homeostasissloughingboundary}
\end{align}
Therefore, in homeostasis, the sloughing boundary decays exponentially in time to balance the growth rate, with decay rate determined by $g(0)$. 

Note that the expression for $L_\mu(t)$ allows us to determine the as-yet unspecified constant $\hat{\gamma}_0$. If we assume that homeostasis began at $t = 0$, then we have the initial condition $L_\mu(0) = 1$. In turn, this fixes the integration constant $\hat{\gamma}_0$ as
\begin{align}
\hat{\gamma}_0 &= -\frac{\hat{v}(l)}{g(0)(1 + \mathcal{S}\hat{n}_\tau(l))}\mathrm{exp}\left(-\int^l_0\frac{g(0) - g(s)}{\hat{v}(s)}ds\right).\label{eq:gamma0constant}
\end{align}
Hence, the expression \eqref{eq:homeostasissloughingboundary} simplifies significantly to
\begin{align}
L_\mu(t) =\mathrm{e}^{-g(0)t}.\label{eq:homeostasissloughingboundarysimplified}
\end{align}

To connect the sloughing boundary $L_\mu$ to a sloughing rate, denote by the quantity $\mu$ the total amount of material sloughed from the rod---physically this would be a measure of the total number of cells extruded from the crypt. This is equivalent to the amount of arc length in the grown configuration that is discounted by mapping $S_0=1$ to the right boundary. Thus, $\mu$ and $\gamma(S_0, t)$ satisfy:
\begin{align}
\mu(t) = S(1) - S(L_\mu(t)) = \int^1_{L_\mu(t)}\gamma dS_0.\label{eq:sloughingrelation}
\end{align}
Differentiating \eqref{eq:sloughingrelation} and applying Leibniz's rule, we find that the sloughing rate, $\dot{\mu}$, relates to $\dot{L}_\mu$ via the integro-differential equation:
\begin{align}
\dot{\mu}  = -\dot{L}_\mu\gamma(L_\mu, t) +\int^{1}_{L_\mu}\dot{\gamma} dS_0.\label{eq:sloughrateeqnhalfinterval}
\end{align} 

By construction, no further growth occurs in the region past $S_0 = L_\mu$. Therefore, we set $\dot{\gamma} = 0$ in the region $S_0\in[L_\mu, 1]$. Furthermore, $\gamma(L_\mu, t) = \hat{\gamma}(l, t)$ and we can write
\begin{align}
\dot{\mu} = -\dot{L}_\mu\hat{\gamma}(l, t).\label{eq:slougheqnsimplified}
\end{align}
Differentiating \eqref{eq:homeostasissloughingboundary} with respect to $t$ and substituting the resulting expression into Equation \eqref{eq:slougheqnsimplified} yields the simplified quantity
\begin{align}
\dot{\mu} = -\frac{\hat{v}(l)}{1 + \mathcal{S}\hat{n}_\tau(l)} =  -\frac{\hat{v}(l)}{1 + \mathcal{S}\hat{n}_x(l)},\label{eq:sloughingratehomeostasis}
\end{align}
where we have used the right boundary condition \eqref{eq:lagrangianBCs} to note that $\hat{n}_\tau(l) = \hat{n}_x(l)$. Therefore, in homeostasis, while $L_\mu(t)$ decays exponentially at a rate determined by the incremental growth at the base, $g(0)$, the sloughing rate $\dot{\mu}$ is constant, with a value that depends on the migration velocity and the stress at the crypt top, $s = l$. Alternatively, the flow velocity equation \eqref{eq:velocityinhomeostasis} can be solved in terms of $\hat{\alpha}(s)$ and $g(s)$, enabling us to express $\dot{\mu}$ as
\begin{align}
\dot{\mu} = \int^l_0\frac{g(s)}{\hat{\alpha}(s)}ds.\label{eq:sloughingratehomeostasisv2}
\end{align}
This expression provides an interpretation of the sloughing rate as balancing the net change of material arc length per unit time, given by integrating the incremental growth over the non-discarded portion of the domain.\footnote{Note that $\frac{ds}{\hat\alpha}=dS$, so that \eqref{eq:sloughingratehomeostasisv2} is equivalent to integrating the incremental growth over the grown `virtual' configuration.} For instance, an increase in growth per unit time means more material must disappear at the boundary to maintain homeostasis. 

\section{Form of mechanical feedback}
\label{sec:formofmechanicalfeedback}
Having formulated the homeostasis framework, we wish to analyse the structure of a homeostatic state. As the homeostatic system is nonlinear, constructing solutions will largely require numerical computation. To proceed, we must first prescribe the functional form of Wnt, $W(s)$, and mechanical feedback, $f(\hat n_\tau-n^*)$. For the Wnt signal, we consider a simple Gaussian form:
\begin{align}
W(s) = \exp\left(-\frac{s^2}{\sigma_W^2}\right),\label{eq:wntgaussian}
\end{align}
which has the desired monotonicity for $s > 0$ but also reflects the rapid decay of Wnt away from the base if the constant $\sigma_W < 1$ \cite{Marshman2002}.
 
Less clear is a reasonable form of mechanical feedback, which is strongly linked to the first question posed in the Introduction: can mechanical feedback account for the bimodal growth profile, given a unimodal Wnt signal? 

To gain insight, we first consider a simplified setting of 1D growth along a line. This will enable us to compare incremental growth profiles for different growth laws without having to account for differing 2D morphologies. Setting $\theta = 0$ and letting $n = n_x$ and $p = p_x$, the Lagrangian governing equations \eqref{eq:lagrangian2Dx}--\eqref{eq:lagrangian2Dmoment} simplify to:
\begin{align}
\frac{\partial s}{\partial S_0} &= \alpha\gamma,\label{eq:1Dgeom}\\
\frac{\partial n}{\partial S_0} &= \alpha\gamma k(s - p), \quad \frac{\partial p}{\partial t} = \rho(s - p),\label{eq:1Dforce}
\end{align}
where $\alpha$ relates to $n$ through the 1D constitutive law
\begin{align}
n = \mathcal{S}^{-1}(\alpha - 1),\label{eq:1Delasticstretch}
\end{align}
and the boundary conditions in a 1D geometry are:
\begin{align}
s(0) = 0, \qquad s(1) = 1.\label{eq:1DBCs}
\end{align}
Consider first a linear form of $f$, so that
\begin{align}
\frac{\dot{\gamma}}{\gamma} = W(s) + \phi(n - n^*),\label{eq:everpresentmechanicalfeedback}
\end{align}
where $\phi$ is a parameter describing sensitivity to mechanical stress, and thus the relative impact of mechanical feedback, and $n^* \le 0$ is the homeostatic stress. The law \eqref{eq:everpresentmechanicalfeedback} models the continual regulation of Wnt signalling due to mechanical feedback. If $n- n^* > 0$, indicating relative tension, incremental growth $\dot{\gamma}\gamma^{-1}$ is increased, while if $n - n^* < 0$, indicating relative compression, $\dot{\gamma}\gamma^{-1}$ is decreased.

\begin{figure*}[t]
	\centering
	\includegraphics[width=\textwidth]{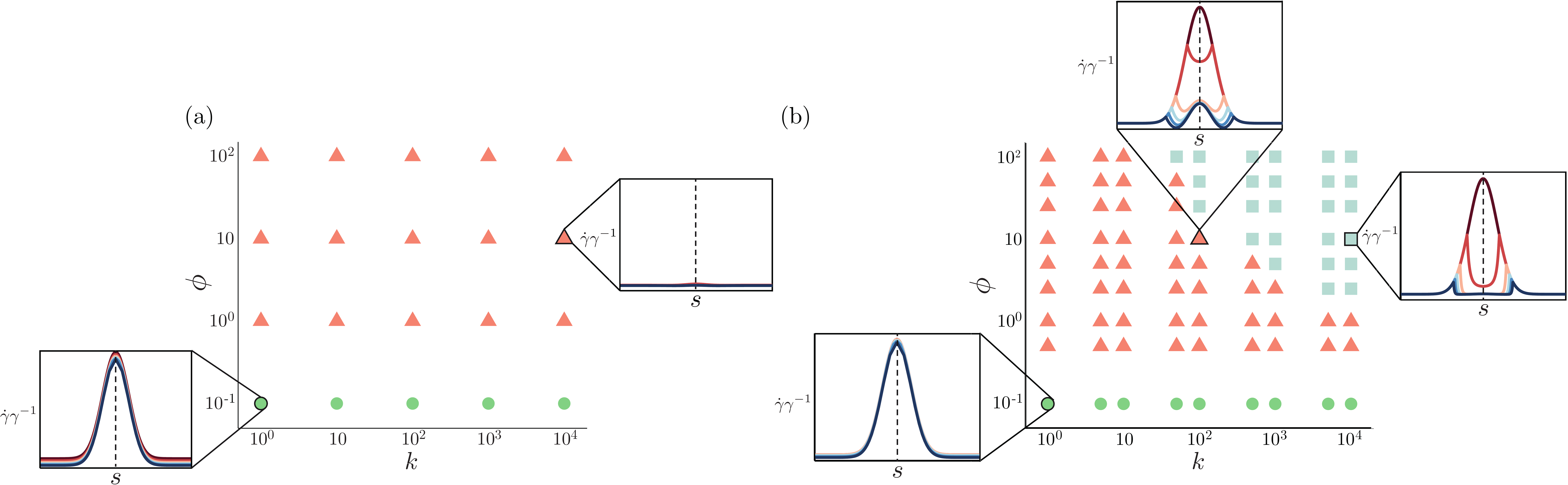}
	\caption{\textbf{Threshold-based mechanical feedback generates a realistic crypt growth structure in 1D.} We have set $\sigma = 0.24$, $\rho = 10$, and $n^* = -0.4$. For selected values of $k$ and $\phi$, we plot the incremental growth profile, $\dot{\gamma}\gamma^{-1}$, at times $t = 0, 1, 2, \dots, 5$. All plots of $\dot{\gamma}\gamma^{-1}$ are to scale. Dark red lines correspond to earlier times, while dark blue lines correspond to later times. \textbf{a} Phase diagram of incremental growth structures for ever-present mechanical feedback \eqref{eq:everpresentmechanicalfeedback}. \textbf{b} Phase diagram of incremental growth structures for threshold-based mechanical feedback \eqref{eq:thresholdbasedmechanicalfeedback}. Threshold-based mechanical feedback generates a richer phase space of growth profiles, including bimodal profiles as observed in the crypt.}
	\label{fig:1Dgrowthphasediagrams}
\end{figure*}

While the form of mechanical feedback considered in the growth law \eqref{eq:everpresentmechanicalfeedback} is often used in studies of mechanical homeostasis \cite{erlich2019homeostatic}, there are two features of this ``ever-present'' feedback that may be interpreted as biologically unrealistic. First, we note that if the threshold stress $n^* < 0$, and the system begins at a stress-free state, $n\equiv0$, then mechanical feedback instantly increases the growth. Second, and more importantly, the growth law \eqref{eq:everpresentmechanicalfeedback} does not actually alter the monotonicity of the spatial profile of $\dot{\gamma}\gamma^{-1}$. In the limiting case of an infinitely stiff foundation ($k\to\infty$), it can be proven (see Appendix \ref{app:everpresentmechanicalfeedbackproof}) that the spatial profile of incremental growth will always have the same monotonicity as $W$. We show that this appears to hold more generally for finite values of $k$ in Fig.\ \ref{fig:1Dgrowthphasediagrams} (discussed below). 

Intuitively, we expect that in order for growth away from the base to overtake growth at the base, where the Wnt signal is highest, we would need the mechanical feedback to be triggered earlier at the base than at other points.
As such, consider a variation where mechanical feedback is triggered only in relative compression:
\begin{align}
\frac{\dot{\gamma}}{\gamma} = W(s) + \phi(n - n^*)\mathrm{H}(n^* - n).\label{eq:thresholdbasedmechanicalfeedback}
\end{align}
Here, the Heaviside function $\mathrm{H}(x)$ effectively generates a spatially-dependent delay of mechanical feedback, as it will now only occur at points $S_0$ for which $n(S_0) < n^*$. 

There are now four parameters in the system, $k$, $\rho$, $\phi$, and $n^*$. Of these, simulations indicate that $k$ and $\phi$ have the strongest qualitative effect on growth profile. The threshold stress $n^*$ has a limited range $-\mathcal{S}^{-1} < n^*$, imposed by the linear constitutive law and the lower bound on stretch $\hat\alpha>0$, and, moreover, $n^*$ should be negative to avoid instantaneously triggering the mechanical feedback (since $n=0$ at $t=0$). The foundation relaxation parameter, $\rho$, only exacerbates or diminishes the effect of $k$ on the spatial variation of $n$, so we do not consider its effect.

To understand the role of the feedback, we thus fix $\mathcal{S} = 1$, $\sigma = 0.24$, $\rho = 10$ and $n^* = -0.4$,and perform a parameter sweep over $k$ and $\phi$. Equations \eqref{eq:1Dgeom}--\eqref{eq:1Dforce} are solved numerically using the MATLAB package \texttt{bvp4c}. For each pair of $(k, \phi)$ values, we simulate Equations \eqref{eq:1Dgeom}--\eqref{eq:1Dforce} subject to one of the growth laws \eqref{eq:everpresentmechanicalfeedback} and \eqref{eq:thresholdbasedmechanicalfeedback} up to $t = 5$ and examine the resulting spatial profile of incremental growth, $\dot{\gamma}\gamma^{-1}$, plotted as a function of $s$ (and including the even extension of the profile to $s<0$). Fig.\ \ref{fig:1Dgrowthphasediagrams} summarises the results of the parameter sweep. 

Fig.\ \ref{fig:1Dgrowthphasediagrams}a characterises the incremental growth in the case of the non-threshold-based growth law \eqref{eq:everpresentmechanicalfeedback}. For all parameter values, the resultant profiles are qualitatively the same, showing a monotonic shape with peak in the middle, albeit with reduced amplitude due to the presence of mechanical feedback. For $(k, \phi) = (10^4, 10)$, the amplitude reduction of $\dot\gamma\gamma^{-1}$ is particularly evident. The monotonic shape is qualitatively the same as the imposed Wnt profile, thus, in this case the mechanical feedback is not sufficient to alter the point of maximal growth.

Fig.\ \ref{fig:1Dgrowthphasediagrams}b considers the threshold growth law \eqref{eq:thresholdbasedmechanicalfeedback}. Here the spatial behaviour of $\dot{\gamma}\gamma^{-1}$ falls into one of three distinct parameter regimes:
\begin{enumerate}
	\item \textbf{Primarily Wnt-driven:} Here, $\dot{\gamma}{\gamma}^{-1}$ is maximal at $s = 0$, and essentially follows $\dot{\gamma}\gamma^{-1} = W(s)$. This occurs when mechanical feedback is insufficient ($\phi$ is too small) to dampen the effect of the Wnt gradient. We note that although our classification of behaviours is based on simulations run up to $t = 5$, the behaviour observed is independent of the simulation end time. We can be certain that this behaviour holds for longer simulation times, since, as stated, the stress is bounded below, $n(S_0, t) > -\mathcal{S}^{-1}$, the inhibition of growth due to mechanical feedback is bounded and if $\phi$ is too small, then the contribution from mechanical feedback will be negligible.
	\item \textbf{Non-monotonic from the base:} In this regime, $\dot{\gamma}\gamma^{-1}$ is still maximal at $s = 0$, but mechanical feedback has reduced growth non-monotonically. There are even parameter regions where $\dot{\gamma}\gamma^{-1} < 0$ for $0 < s < 1$. This behaviour arises for the widest range of parameter values, as neither $k$ nor $\phi$ are high enough to give rise to bimodal behaviour.
	\item \textbf{Maximal away from the base and top:} In this case, $\dot{\gamma}\gamma^{-1}$ no longer attains a local maximum at $s = 0$. As such, its even extension is bimodal. This occurs when mechanical stress, $n(S_0, t)$, exhibits sufficient spatial variation due to growth ($k$ is sufficiently large) and mechanical feedback is sufficiently strong ($\phi$ is sufficiently large) such that mechanical inhibition occurs in the base first, before other regions are affected.  If $k$ is small, then \eqref{eq:1Dforce} implies that the rod stress $n(S_0, t)$ is effectively constant. Contrastingly, if we take $k \to \infty$, then $n(S_0, t) = \mathcal{S}^{-1}(\gamma^{-1} - 1)$, which is minimal where growth $\gamma$ is maximal, and vice versa.
\end{enumerate}
This analysis demonstrates that mechanical feedback, together with unimodal biochemical signalling, can produce a bimodal form of incremental growth, if the feedback has threshold dependence and sufficiently high values of $k$ and $\phi$. That is, given a background biochemical signal that decreases monotonically from the crypt base to the top, if growth is regulated by mechanical stress, but is only triggered at sufficient compression, then with strong feedback and high resistance to deformation from the underlying substrate, the growth profile can be qualitatively altered to no longer be monotonic. It remains to demonstrate that this result carries over to the 2D morphology; for this, we turn to solving the full homeostatic system. 

\section{Computing homeostasis}
\label{sec:computinghomeostasis}
To examine properties of homeostasis, in this section we solve the homeostatic BVP for the spatial variables $\mathbf{U}$ numerically, extract the growth profile and sloughing rate, and examine the properties and dependence on model parameters. We will also analyse the \emph{dynamic stability} of the homeostatic solutions, which relates to the robustness of the homeostasis profiles while maintaining quasi-static equilibrium, and discuss how this compares to the more classical \emph{inertial stability}. 
\begin{figure*}[t!]
	\centering
	\includegraphics[width=0.9\textwidth]{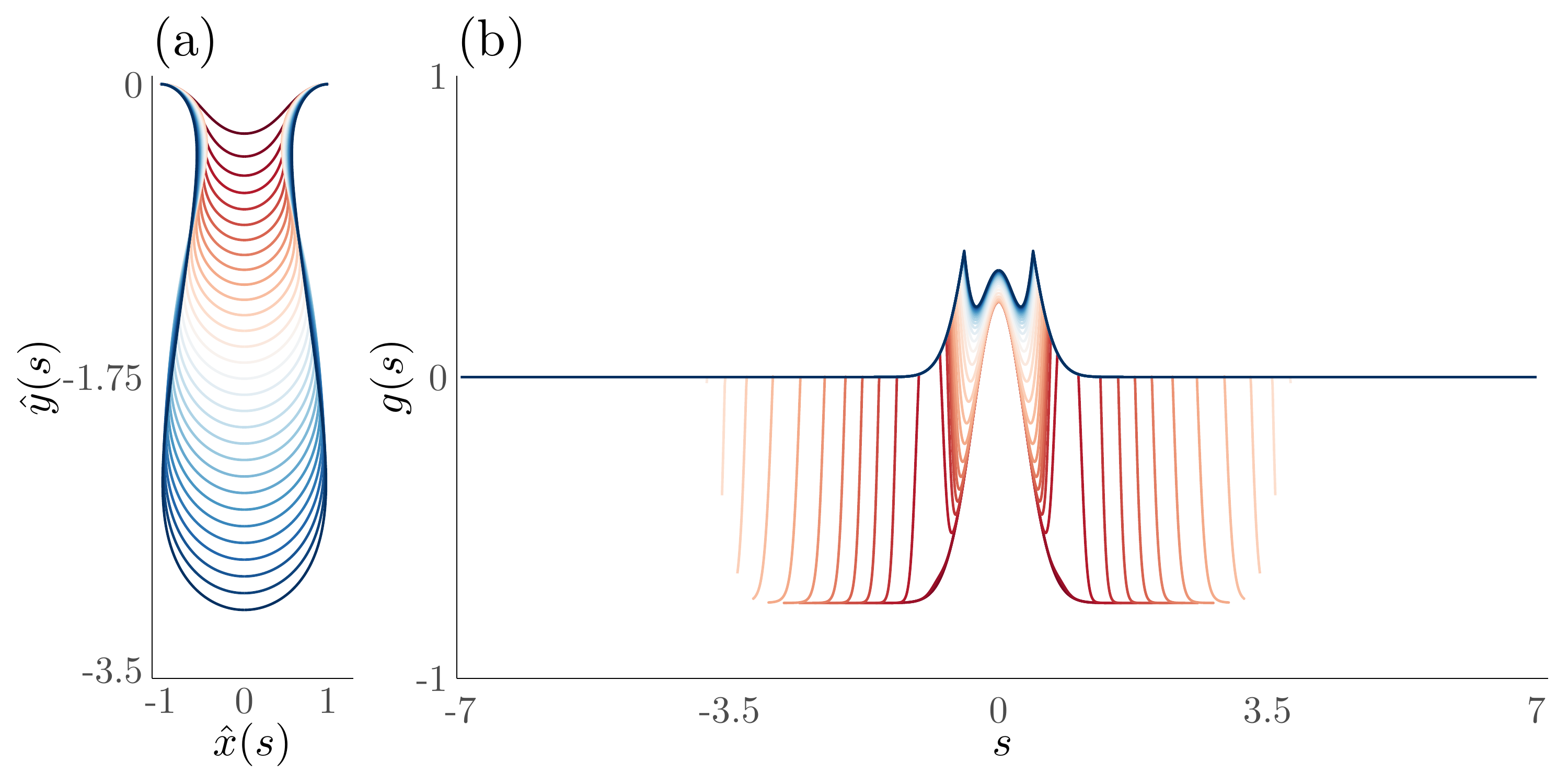}
	\caption{\textbf{Computed homeostatic solutions for different values of total rod length, $l$.} Solutions to Equations \eqref{eq:eulerian2Dx}--\eqref{eq:S0compatibility} are obtained from numerical continuation in $l$ over the values $l = 1.2, 1.4, \dots, 6.8, 7$. Dark red lines indicate solutions for smaller values of $l$, while dark blue lines indicate solutions for larger values of $l$. The chosen model parameter values are $k = 868.056\  (k_f = 0.01)$, $\rho = 10$, $\phi = 0.75$, and $\hat{n}_\tau^* = -3$. \textbf{a} The homeostatic morphologies $(\hat{x}(s), \hat{y}(s))$ for increasing $l$. Each solution has been reflected about $x = 0$ to better represent possible crypt morphologies. \textbf{b} The resulting homeostatic incremental growth profiles $g(s)$ for increasing $l$, which have been reflected about $s = 0$ to better reflect possible proliferative structures of the crypt. For larger values of $l$, the spatial structure of $g(s)$ resembles the crypt's proliferative structure, demonstrating the link between morphology and homeostatic growth.}
	\label{fig:2Dhomeostaticmorphologies}
\end{figure*}

In the Eulerian framework, note that the total arc length $s=l$ is effectively a free parameter; that is, for any given $l$, one could seek a solution to the BVP. If we model the current arc length to be only slightly larger than the initial domain length, i.e. $l=1+\epsilon $ for $\epsilon\ll1$, then the crypt shape will be nearly flat. In this limit, we can thus determine a solution as an asymptotic expansion about the state $\hat\theta\equiv0$. Similar to a buckling analysis, infinitely many solutions exist, with increasing mode, yet only one will be stable in the classical sense. Having computed the stable mode, we then perform a numerical continuation with increasing $l$, thus producing more crypt-like homeostatic profiles with deeper invaginations. 

\subsection{Examining the homeostatic solutions}
\label{sec:homeostasiscontinuations}
For 2D morphologies, we consider an altered form of threshold-dependent mechanical feedback:
\begin{align}
g(s) = W(s) + \phi\tanh(\hat{n}_\tau - \hat{n}_\tau^*)\mathrm{H}(\hat{n}_\tau^* - \hat{n}_\tau),\label{eq:saturatingmechanicalfeedback}
\end{align}
where the effect of mechanical feedback saturates. Not only is this form of mechanical feedback more physically realistic, but it is more amenable to numerical continuation in $l$. For the form of mechanical feedback specified by \eqref{eq:thresholdbasedmechanicalfeedback}, setting $\phi$ large enough to generate both the desired homeostatic morphology and, consequently, the correct incremental growth profile for larger values of $l$ can result in numerical singularities for smaller values of $l$ if $g(0) \le 0$. As we demonstrate in Appendix B, the altered form \eqref{eq:saturatingmechanicalfeedback} has the same qualitative features as \eqref{eq:thresholdbasedmechanicalfeedback} in the 1D geometry.

For given $l$, the time-independent variables $\mathbf{U}$ are found as solutions to the Eulerian system \eqref{eq:eulerian2Dx}--\eqref{eq:S0compatibility}, using the Eulerian boundary conditions \eqref{eq:homeostasisbcs}. These were computed numerically using a shooting method implemented with the Mathematica package \texttt{NDSolve}. 
We set the foundation stiffness parameter, $k = 868.056$, corresponding to a foundation stiffness scaling of $k_f = 0.01$, generating a mode one instability upon buckling; the foundation relaxation parameter to $\rho = 10$, corresponding to a rapidly-relaxing foundation that contributes to a deeply-invaginated morphology; the width of the Gaussian Wnt profile to $\sigma_W = 0.24$; the mechanical feedback strength to $\phi = 0.75$ in order to sufficiently alter the monotonicity of the homeostatic incremental growth $g(s; \hat{n}_\tau)$ for larger values of $l$; and the threshold axial stress to $\hat{n}_\tau^* = - 3$ so that mechanical feedback inhibits growth in the crypt base ($s = 0$) but not at the crypt top ($s = l$).

In Fig.\ \ref{fig:2Dhomeostaticmorphologies}, we plot the homeostatic morphologies, $(\hat{x}(s), \hat{y}(s))$, and resultant homeostatic incremental growth profiles, $g(s)$, that were obtained from numerical continuation in $l$---each colour corresponds to a value of $l$ increasing from $l=1.2$ (red) to $l=7$ (blue). Unsurprisingly, in Fig.\ \ref{fig:2Dhomeostaticmorphologies}a, as the total rod length increases, the rod morphology becomes increasingly invaginated, a known feature of post-buckled elastic rods \cite{Edwards2007, Nelson2011, Almet2019, Chirat2013}.  Fig.\ \ref{fig:2Dhomeostaticmorphologies}b shows that the incremental growth profile also varies with $l$. For smaller values of $l$, where the morphologies exhibit little invagination, the growth profile is in fact non-negative only around $s = 0$, indicating that the homeostatic profile at small lengths can only be maintained with a resorption of material over most of the domain. Interestingly, we observe a qualitative transition in the growth profile, so that at larger values of $l$, when the morphology is significantly invaginated, the incremental growth profiles are both non-negative everywhere and also maximal away from the base, resembling the bimodal proliferative structure of the crypt. This confirms the intuition drawn from a 1D geometry in Sect.\ \ref{sec:formofmechanicalfeedback} on producing bimodal growth through mechanical feedback, while the qualitative dependence on $l$ highlights the non-trivial relationship between morphology and growth structure. 
\begin{figure*}[t!]
	\centering
	\includegraphics[width=0.95\textwidth]{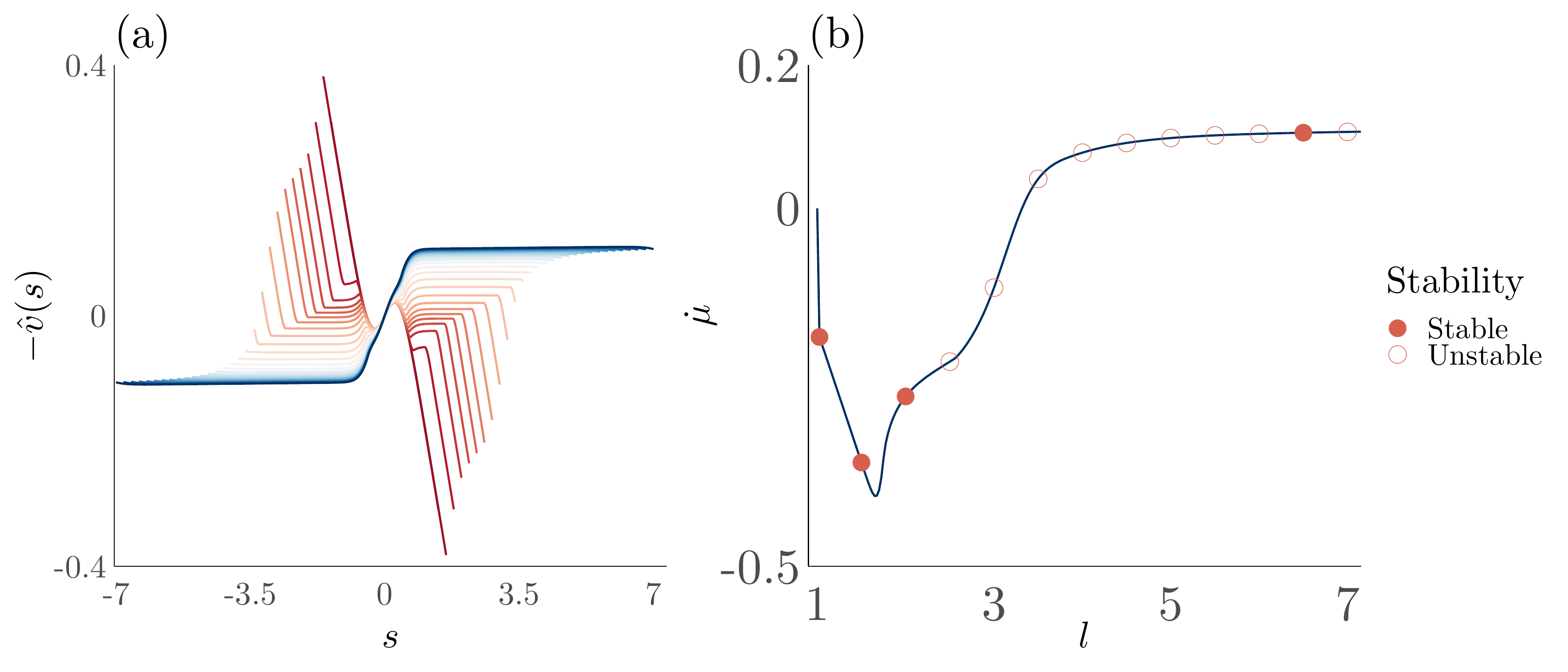}
	\caption{\textbf{Biologically-relevant measurements from the homeostatic solutions.} \textbf{a} The migration velocity, defined by $-\hat{v}(s)$. Migration velocity solutions are obtained from numerical continuation of Equations \eqref{eq:eulerian2Dx}--\eqref{eq:S0compatibility} in $l$, over the range $l = 1.2, 1.4, \dots, 6.8, 7$, for assigned model parameter values of $k = 868.056\ (k_f = 0.01)$, $\rho = 10$, $\phi = 0.75$, and $\hat{n}_\tau^* = -3$. Dark red lines correspond to solutions for smaller values of $l$, while dark blue lines correspond to solutions for larger values of $l$. In line with the emergence of realistic $g(s)$ profiles, physically realistic migration velocity profiles only begin to emerge for larger values of $l$. \textbf{b} The sloughing rates $\dot{\mu}$ for various values of $l$, which balance incremental growth to dynamic tissue homeostasis. We have marked the dynamic stability of several solutions along this curve at $l = 1.025, 1.5, 2, \dots, 6.5, 7$.}
	\label{fig:2Dhomeostaticmeasurements}
\end{figure*}

In addition to the homeostatic growth structure, in Fig.\ \ref{fig:2Dhomeostaticmeasurements} we plot the migration velocity profile $-\hat{v}(s)$ and sloughing rate $\dot\mu$ for increasing length $l$. These parameters form continuum versions of experimentally-accessible measurements for the crypt, and with significant biological relevance, as they provide insight into the health status of the epithelium, where efficient migration and turnover are important for maintaining the intestinal epithelium \cite{potten1997regulation, sansom2004loss}. Fig.\ \ref{fig:2Dhomeostaticmeasurements}a shows that for smaller values of $l$, when there is little crypt invagination, the migration velocity is negative over much of the domain (for $s>0$), akin to cells migrating \emph{down} into the crypt, towards the base; though the velocity remains positive near the base, implying a stagnation point. It is only at larger values of $l$ that the velocity attains a more realistic profile, where it represents cells migrating purely outward. In particular, there is an approximately-linear increase in velocity from $s = 0$, corresponding to the region of growth $g(s)>0$, and a flat velocity in the outer part of the domain where $g(s)\approx0$. This feature is consistent with experimental observations that also report a linear increase in migration velocity from the crypt base \cite{kaur1986cell, krndija2019active}. 

For each length $l$, the sloughing rate $\dot\mu$ needed to maintain the homeostatic state is plotted in Fig.\ \ref{fig:2Dhomeostaticmeasurements}b. Note that a positive value of $\dot\mu$ corresponds to a net loss of material (extrusion of cells), while $\dot\mu<0$ would require the unphysical addition of material at the boundary. We observe that the sloughing is only positive for larger $l$, when the morphology is sufficiently invaginated. In particular, note that for larger values of $l$, $\dot{\mu} \approx 0.1$, which corresponds to a net turnover time in the crypt of $\dot{\mu}^{-1}$, which in dimensional units, corresponds to $T\dot{\mu}^{-1} \approx 10$ days and on the order of turnover time scales observed \emph{in vivo} \cite{gehart2018tales}.

We emphasise that the solutions presented in Fig.\ \ref{fig:2Dhomeostaticmorphologies}--\ref{fig:2Dhomeostaticmeasurements}, obtained from numerical continuation in $l$, are not to be interpreted as a time sequence; that is, they do not represent a transition from development to homeostasis. Rather, homeostasis is \emph{imposed} at each length, and time is a hidden parameter at each length through which total growth $\hat\gamma$ and initial arc-length $\hat S_0$ continue to evolve. 

The quantities depicted in Figs \ref{fig:2Dhomeostaticmorphologies} and \ref{fig:2Dhomeostaticmeasurements} reflect a delicate balance between growth, stress, and morphology in homeostasis. These plots also present two qualitatively distinct regimes: homeostasis at small $l$ is characterised by very little invagination and a monotonic growth profile with resorption and a downwards migration of material over much of the domain. This state consists of material lost from negative incremental growth that can only be maintained by ``negative'' sloughing at the boundary, which adds material at the boundary to balance negative growth. These characteristics are all highly unphysical for the crypt. On the other hand, at larger $l$ the picture is completely different: here, the homeostatic morphology has a deep invagination, bimodal growth profile with positive (or zero) growth throughout and an outwards migration of material, where the net gain of material is maintained by positive sloughing at the boundary. Each of these features is consistent with observations in the crypt. Indeed, the differences between the two regimes may provide some insight as to why the crypt has such a deeply-invaginated morphology in homeostasis, or rather, why homeostasis is not established until the crypt is deeply invaginated. The characteristics needed for homeostasis at small $l$, in particular negative growth and sloughing, are not physically possible i nthis model, so the system cannot remain in such a state.

\subsection{Dynamic stability analysis}
\label{sec:stability}
An important feature of the intestinal crypt and of biological homeostasis in general, is its robustness over time and the ability of the system to return to homeostasis after perturbation due to, for example, injury or genetic mutation. Having computed the structure of typical homeostatic solutions and demonstrated that properties of the crypt can be replicated within our continuum framework, we now investigate the robustness of these homeostatic states to perturbations. 
In other words, we are concerned with the notion of \emph{dynamic} stability: whether a homeostatic state is recovered following a perturbation to the system variables. Here we restrict to small dynamic perturbations, as they are amenable to a linear stability analysis. 

Note that dynamical stability is not mechanical stability in the classical sense of incorporating inertia within the momentum balance equations. Rather, the question is whether the system returns to the homeostatic state if the variables are perturbed on the slow timescale of growth \cite{erlich2019homeostatic}. To answer this question, we perform a linear stability analysis on the quasi-static system\footnote{Since we follow only the critical buckling mode, inertial stability is already established for the solutions under consideration.}. Here we must take care to perturb the system in a consistent way, given that there are two time-dependent variables, $\hat{S}_0(s, t)$ and $\hat{\gamma}(s, t)$, while the other dependent variables, summarised in \eqref{eq:timeindependentsolns}, are functions of $s$ only. 

We perturb the time-independent variables as
\begin{align}
\hat{\nu}(s) = \hat{\nu}^{(0)}(s) + \varepsilon\hat{\nu}^{(1)}(s)\mathrm{e}^{\sigma t},\label{eq:timeindeptsolnpert}
\end{align}
where $\hat{\nu}(s) \in\mathbf{U}$; $\varepsilon \ll 1$ is an arbitrarily small parameter; and the sign of $\sigma \in \mathbb{R}$ determines the growth or decay of the perturbation and, consequently, the stability of the homeostatic solutions. Here and below, the superscript $(0)$ refers to the homeostatic profile, such as those obtained in Sect.\ \ref{sec:homeostasiscontinuations}, while superscript $(1)$ denotes the perturbed spatial profile, which must be determined as part of the stability analysis.

In Sect.\ \ref{sec:growthsloughinghomeostasis}, we showed that $\hat{S}_0(s, t)$ and $\hat{\gamma}(s, t)$ decay and grow, respectively, on a timescale determined by $g(0)$, the homeostatic incremental growth at the base. In order to expand the time-dependent variables in a manner that is consistent with the time-independent variables, we write
\begin{align}
&\hat{S}_0(s, t) = \Sigma^{(0)}(s)\mathrm{e}^{-\beta t} + \varepsilon \Sigma^{(1)}(s)\mathrm{e}^{-\beta t}\mathrm{e}^{\sigma t},\label{eq:S0pert}\\
&\hat{\gamma}(s, t) = \hat{\gamma}_0\Gamma^{(0)}(s)\mathrm{e}^{\beta t} + \varepsilon \hat{\gamma}_0\Gamma^{(1)}(s)\mathrm{e}^{\beta t}\mathrm{e}^{\sigma t},\label{eq:gammapert}
\end{align}
where $\beta = g^{(0)}(0)$. 
The system at $O(1)$ is satisfied by the homeostatic solution. At $O(\varepsilon)$, we obtain a linearised BVP with eigenvalue $\sigma$. We have solved this by implementing a determinant method that involves integrating multiple linearly-independent copies of the system from $s=0$ to $s=l$; the details are presented in Appendix \ref{app:stabilityanalysis}. For the homeostatic solution to be dynamically stable, all eigenvalues $\sigma$ must satisfy $\mathrm{Re}(\sigma) < 0$. If there is at least one eigenvalue $\mathrm{Re}(\sigma) > 0$, then the solution is dynamically unstable.

We have computed the stability of homeostatic solutions for various values of $l$ in Fig.\ \ref{fig:2Dhomeostaticmeasurements}, and have labelled the stability in Fig.\ \ref{fig:2Dhomeostaticmeasurements}b. We find that for smaller $l$, up to $l = 2$, the homeostatic state is stable. As $l$ increases, a transition to unstable homeostatic states is observed. A lone exception of the solutions we examined is at $l = 6.5$, for which the solution is computed to be stable. It is an artefact of our chosen discretisation in $l$ that the only stable solution that is found for larger values of $l$ is at $l = 6.5$; it is much more likely that there is a small region about $l = 6.5$ for which all solutions are stable. Also of note is that we find (results not plotted) that higher buckling modes, which are inertially unstable, are also dynamically stable for small values of $l$, a feature that highlights the distinction between dynamic and inertial stability. 

As stated before, it is only at larger $l$ that the homeostatic state resembles the characteristics of the crypt. While we have only presented results for a small subset of parameter space, and it would be imprudent to reach too strong of conclusions in the context of the crypt, it is nevertheless interesting that we have uncovered a single homeostatic state that has both physically-realistic characteristics and is dynamically stable. How a biological system regulates its size and {\it selects} homeostasis is one of the fundamental open questions in biology \cite{Goriely:2016tc}; while our model is undoubtedly too simplified to answer this question for the crypt, the ideas presented here may provide new insight as well as a new tool of analysis.

\section{Discussion}
\label{sec:discussion}
In this paper, we developed a mathematical model of tissue homeostasis in the intestinal crypt. The model built on morphoelastic rod theory, a continuum mechanics framework. 
Modelling tissue homeostasis as a dynamic process required careful translation of the mechanical description from the typical Lagrangian frame to the Eulerian frame, where the concept of tissue homeostasis is most naturally defined. A continuum framework enabled us to investigate the role of mechanics both in generating the proliferative structure of the crypt and maintaining this structure in homeostasis.

A key starting point for our framework was a clear definition of homeostasis. Here we translated the biologist's view---a spatially-heterogeneous treadmilling state with constant growth and velocity at each point in a lab frame---to our setting, and this produced a clear statement of certain system variables being time-independent when expressed in an Eulerian frame. 
In doing so, we demonstrated how the same quantities in the Lagrangian reference frame must evolve over time to maintain a static Eulerian frame, resulting in a migration of (Lagrangian) material points out of the crypt, not unlike the conveyor belt mechanism observed \emph{in vivo} \cite{krndija2019active}. Correspondingly, the removal of material in the form of sloughing emerges naturally to balance growth in homeostasis. A clear advantage of the continuum framework was that it enabled closed-form solutions for both growth and the sloughing rate to be constructed.

In order to simulate growth and homeostasis in the crypt, we required explicit assumptions about the key contributors to growth. Here, we assumed mechanochemical growth, focussing on the role of mechanical (axial) stress in regulating the well-known Wnt signal profile that is present along the crypt. We showed that threshold-dependent mechanical feedback, where mechanical inhibition of growth is only triggered at points that are sufficiently compressed, can generate the observed growth structure of the crypt. This is not unlike the contact inhibition model that is considered in individual-based crypt models \cite{Osborne2017}. In contrast, the commonly used ``ever-present'' mechanical feedback law was incapable of generating the correct growth structure. 

In homeostasis, the time-independent variables in the Eulerian system decoupled from the time-dependent variables, allowing the homeostatic state to be fully resolved by solving a spatial BVP. Therefore, we were able to compute homeostatic states using numerical continuation, much like standard buckling problems. Here, the continuation parameter was the total rod length in the homeostatic state. Numerical solutions revealed how the homeostatic incremental growth structure, migration velocity, and sloughing rate depends on the morphology (and stress profiles), suggesting that a significantly-invaginated morphology may actually be necessary for crypt homeostasis, as plausible homeostatic growth and velocity profiles and sloughing rates emerged only for deep invaginations. However, dynamic stability analysis, which provided insight into the ``robustness'' of the constructed homeostatic solutions, revealed that many of these homeostatic states that generated the correct growth structure were dynamically unstable, despite being inertially stable, i.e. the preferred buckling mode. From an experimental point of view, this type of instability would correspond to a perturbation of the homeostatic state (say through injury or a small change in growth rate) causing a significant change in growth structure and/or morphology.

Extensions of the work presented here may naturally proceed in two distinct directions: specialisation to the crypt, and generalisation to explore more broadly the role of mechanics in homeostasis.
With regards to the former, there are a number of ways to specialise our framework to that of the crypt. One assumption we have made is that all material parameters are spatially and temporally homogeneous. Viewing points along the rod as representative of a growing line of cells along the crypt is akin to the assumption that all cells have the same physical properties and response to biochemical signalling at all times. Spatial heterogeneity was only included in the Wnt signal profile, which consequently induced a heterogeneous response to mechanical feedback. It would be straightforward to include spatial or temporal dependence on other system properties, though of course one may have to trade analytical tractability for physical accuracy. 

Also of note is that mechanical feedback was modelled as a phenomenological process that occur instantaneously when triggered. However, it is known that each crypt contains a diverse population of cell types with varying mechanical properties and responses to chemical signals (such as Wnt) \cite{spit2018tales}. In particular, it has been shown that YAP and TAZ, known mechanotransduction pathways, regulate the cellular response to Wnt signalling due to mechanical stress \cite{azzolin2014yap}. At the cellular scale, Wnt is regulated by YAP and TAZ shuttling between the cell nucleus and cell cytoplasm, depending on mechanical stress. The shuttling mechanism means that there is a certain time lag between the activation of mechanical feedback and the resultant inhibition of Wnt. In other words, the response to mechanical feedback, modelled through the parameter $\phi$, may in fact be time-dependent. Connecting model parameters, such as the Wnt response, $W(s)$, and mechanical feedback strength, $\phi$, to more detailed mathematical models of biochemical signalling pathways, such as Wnt and YAP/TAZ, which can be more readily perturbed and tested in wet lab experiments, would allow us to validate and refine the role of mechanics in the crypt. 

In order to keep the biochemistry as simple as possible, here we modelled Wnt as the sole biochemical regulator of proliferative capacity and investigated possible ways that Wnt may be regulated by mechanical stress. By modelling mechanochemical growth, we showed that it is possible to generate the proliferative structure of the crypt through this minimal growth law. Additionally, the assumption of mechanochemical growth increased the analytical tractability of the homeostasis framework considerably. However, there are numerous signalling pathways involved in crypt homeostasis \cite{spit2018tales}. For example, BMP signalling has been established to negative regulate proliferation in the crypt by driving terminal differentiation of stem cells \cite{haramis2004novo}. As such, it may be possible to generate the same proliferative structure through purely biochemical processes, say, through the interaction of Wnt and BMP signalling. 

In the other direction, separate from the crypt, a broader goal of this paper was to formulate a continuum framework that links growth and mechanics in dynamic homeostasis. Here, we have uncovered a number of interesting features and challenges that may naturally be explored in more detail and perhaps in a more generic setting. For one, an intriguing connection exists between the homeostatic sloughing we defined at the boundary and studies that have modelled surface growth as an evolving reference configuration \cite{zurlo2017printing, zurlo2018inelastic, truskinovsky2019nonlinear}. While we have studied a 1D rod structure, both for simplicity and as a reasonable idealisation of the crypt, in principle, the ideas we have presented for constructing homeostasis in an Eulerian frame could carry over to 2D surfaces and 3D bodies, though almost certainly with added complications.
Also of interest are several papers \cite{tomassetti2016steady, abi2019kinetics, abi2020surface, abeyaratne2020treadmilling} that have considered how surface growth can lead to ``treadmilling'' and the notion of a ``universal growth path'', a concept that is connected to the developmental trajectory towards homeostasis, a feature that is absent from our study. Computing such trajectories within our modelling framework would certainly provide valuable insight, though it is not clear how our framework would have to be adapted to do so. While this forms an appealing avenue of future work, the nonlinear behaviour and rich solution structure we have uncovered in the case of imposed homeostasis highlights the significant challenge in providing an answer to the fundamental question of how growth is determined and regulated in biology.

\section*{Acknowledgments}
This work was supported by Cancer Research UK (CRUK) grant number C5255/A23225, through a Cancer Research UK Oxford Centre Prize DPhil Studentship. 

\bibliographystyle{spmpsci} 
\bibliography{CryptHomeostasis}

\appendix
\section{Ever-present mechanical feedback does not disrupt a monotonic Wnt signal}
\label{app:everpresentmechanicalfeedbackproof}
Here, we prove that in the limiting case of an infinitely-stiff foundation, $k \to \infty$, on a 1D geometry, the type of ever-present mechanical feedback modelled in \eqref{eq:everpresentmechanicalfeedback},
\begin{align*}
\frac{\dot{\gamma}}{\gamma} = W(s) + \phi(n - n^*),
\end{align*}
does not alter the monotonicity of the Wnt signal, $W(s)$, and thus the monotonicity of incremental growth, $\dot{\gamma}\gamma^{-1}$. We deduce that for an infinitely-stiff foundation, ever-present mechanical feedback will not generate a bimodal proliferative structure. Combined with the results from Fig.\ \ref{fig:1Dgrowthphasediagrams}a, which are for a finite foundation stiffness, we conclude that this simple form of ever-present mechanical feedback is insufficient for replicating the proliferative structure of the crypt. While it is more difficult to prove that more complex forms of ever-present mechanical feedback cannot generate bimodal proliferative structures, we conjecture that threshold-dependent mechanical feedback is necessary to generate a bimodal proliferative structure, as well as being a more biologically realistic form of mechanical feedback. 

In the limiting case of an infinitely-stiff foundation, $k \to \infty$, we can solve Equations \eqref{eq:1Dgeom}--\eqref{eq:1DBCs} analytically and obtain solutions for the current arc length, $s(S_0, t)$, and stress, $n(S_0, t)$:
\begin{align}
s(S_0, t) = S_0, \qquad n(S_0, t) = \frac{1 - \gamma}{\mathcal{S}\gamma}.\label{eq:1Dinfinitestiffnesssoln}
\end{align}
Substitution of the solution \eqref{eq:1Dinfinitestiffnesssoln} into the growth law \eqref{eq:everpresentmechanicalfeedback} yields a first-order, linear ODE for $\gamma(S_0, t)$:
\begin{align}
\dot{\gamma} = \gamma W(S_0) + \phi\left(\mathcal{S}^{-1}(1 - \gamma) - n^*\gamma\right).
\end{align}
Recalling the initial condition $\gamma(S_0, 0) = 1$, we deduce the following expressions for $\gamma$ and $\dot{\gamma}\gamma^{-1}$:
\begin{align}
\gamma(S_0, t) &= \frac{A_0 + A_1\mathrm{e}^{-gt}}{g}, \label{eq:gammasolninfinitestiffness}\\
\frac{\dot{\gamma}}{\gamma} &= \frac{-A_1g\mathrm{e}^{-gt}}{A_0 + A_1\mathrm{e}^{-gt}}\label{eq:incgammasolinfinitestiffness}
\end{align}
where $A_0$, $A_1$, and $g$ are given by:
\begin{align}
A_0 &= \phi\mathcal{S}^{-1},\nonumber\\
A_1 &= \phi n^* - W(S_0),\nonumber\\
g &= \phi(\mathcal{S}^{-1} + n^*) - W(S_0).\label{eq:A0A1g}
\end{align}
Note that $A_0$, $A_1$, and $g$ are all functions of $W(S_0)$.
With the closed-form expression \eqref{eq:incgammasolinfinitestiffness}, we can now make the following claim.
\paragraph{Claim:} Let $\dot{\gamma}\gamma^{-1}$ evolve according to \eqref{eq:incgammasolinfinitestiffness}. Assume that $W(S_0)$ is a monotonically decreasing function of $S_0$. Then $\dot{\gamma}\gamma^{-1}$ is also a monotonically decreasing function of $S_0$.
\paragraph{Proof:} We must show that for $S_1 \ge S_2$, the following inequality is satisfied for all time $t > 0$:
\begin{align}
\frac{\dot{\gamma}}{\gamma}\Big |_{S_0 = S_1} \ge \frac{\dot{\gamma}}{\gamma}\Big |_{S_0 = S_2}.\label{eq:incgammainequality}
\end{align}
Denote $w_1 := W(S_1)$ and $w_2 := W(S_2)$, respectively. Substituting \eqref{eq:A0A1g} into expression \eqref{eq:incgammasolinfinitestiffness}, we must show:
\begin{align}
&\frac{(w_1 - \phi n^*)(\phi(\mathcal{S}^{-1} + n^*) - w_1)\mathrm{e}^{-(\phi(\mathcal{S}^{-1} + n^*) - w_1)t}}{\phi\mathcal{S}^{-1} + (\phi n^* - w_1))\mathrm{e}^{-(\phi(\mathcal{S}^{-1} + n^*) - w_1)t}}\nonumber\\
&\ge \frac{(w_2 - \phi n^*)(\phi(\mathcal{S}^{-1} + n^*) - w_2)\mathrm{e}^{-(\phi(\mathcal{S}^{-1} + n^*) - w_2)t}}{\phi\mathcal{S}^{-1} + (\phi n^* - w_2)\mathrm{e}^{-(\phi(\mathcal{S}^{-1} + n^*) - w_2)t}}.\label{eq:incgammainequalityw1w2}
\end{align}
By our assumption, $w_1 \ge w_2$. It then follows that $w_2 - \phi n^* \ge w_1 - \phi n^*$ and $\mathrm{e}^{-(\phi(\mathcal{S}^{-1} + n^*) - w_2)t} \ge \mathrm{e}^{-(\phi(\mathcal{S}^{-1} + n^*) - w_1)t}$. Therefore, we may bound the right-hand side of the inequality \eqref{eq:incgammainequalityw1w2} above by:
\begin{align}
&\frac{(w_2 - \phi n^*)(\phi(\mathcal{S}^{-1} + n^*) - w_2)\mathrm{e}^{-(\phi(\mathcal{S}^{-1} + n^*) - w_2)t}}{\phi\mathcal{S}^{-1} + (\phi n^* - w_2)\mathrm{e}^{-(\phi(\mathcal{S}^{-1} + n^*) - w_2)t}}.\nonumber\\
&\le  \frac{(w_1- \phi n^*)(\phi(\mathcal{S}^{-1} + n^*) - w_2)\mathrm{e}^{-(\phi(\mathcal{S}^{-1} + n^*) - w_2)t}}{\phi\mathcal{S}^{-1} + (\phi n^* - w_1)\mathrm{e}^{-(\phi(\mathcal{S}^{-1} + n^*) - w_1)t}}.
\end{align}
Therefore, to prove that the inequality \eqref{eq:incgammainequalityw1w2} holds, it suffices to show that:
\begin{align}
&(\phi(\mathcal{S}^{-1} + n^*) - w_2)\mathrm{e}^{-(\phi(\mathcal{S}^{-1} + n^*) - w_2)t}\nonumber
\\&\le  (\phi(\mathcal{S}^{-1} + n^*) - w_1)\mathrm{e}^{-(\phi(\mathcal{S}^{-1} + n^*) - w_1)t}.\label{eq:incgammainequalityw1w2simplified}
\end{align}
Equivalently, we may simplify \eqref{eq:incgammainequalityw1w2simplified} thusly and show:
\begin{align}
\mathrm{e}^{(w_1 - w_2)t} \ge \frac{w_2 - \phi(\mathcal{S}^{-1} + n^*)}{w_1 - \phi(\mathcal{S}^{-1} + n^*)}.\label{eq:incgammainequalityexpinequality}
\end{align}
Observe that, as $w_1 \ge w_2$, the right-hand side of \eqref{eq:incgammainequalityexpinequality} is bounded above by one. Moreover, as $t > 0$, it must also be true that $\mathrm{e}^{(w_1 - w_2)t} \ge 1$ for all $t > 0$. Therefore, the inequality \eqref{eq:incgammainequalityexpinequality} and, consequently, the inequality \eqref{eq:incgammainequality} hold for all time $t > 0$. That is, for a monotonically decreasing Wnt signal profile, the resultant incremental growth for the ever-present mechanical feedback law \eqref{eq:everpresentmechanicalfeedback} is also monotonically decreasing function of $S_0$. Hence, on an infinitely-stiff foundation, ever-present mechanical feedback is insufficient to generate the correct proliferative structure.

\begin{figure*}[t!]
	\centering
	\includegraphics[width=\textwidth]{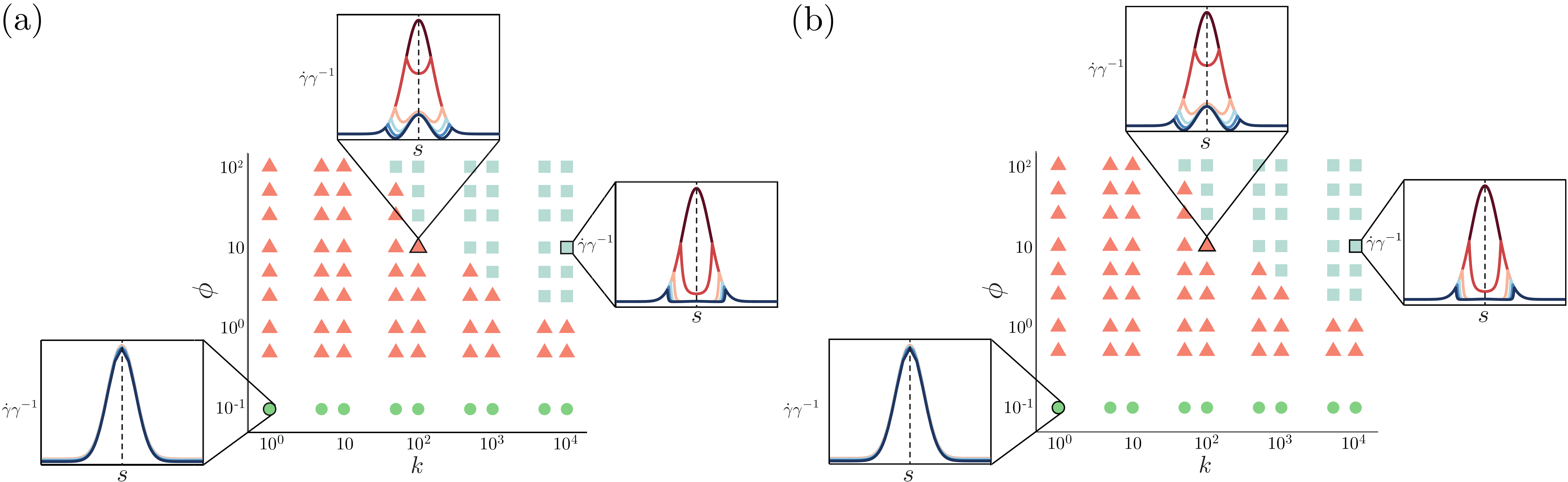}
	\caption{\textbf{Saturating threshold-dependent mechanical feedback is indistinguishable from linear threshold-dependent mechanical feedback in a 1D geometry.} \textbf{a} The phase space of growth profiles for linear threshold-dependent mechanical feedback \eqref{eq:thresholdbasedmechanicalfeedback}, as shown in Fig.\ \ref{fig:1Dgrowthphasediagrams}b. \textbf{b} The phase diagram for saturating threshold-dependent mechanical feedback \eqref{eq:1Dsaturatingmechanicalfeedback}. The phase diagrams and representative incremental growth profiles are identical in a 1D geometry. }
	\label{fig:1Dlinearvssaturatingthresholdbasedfeedback}
\end{figure*}

\section{Saturating threshold-dependent mechanical feedback in 1D}
\label{app:saturatingvslinearthreshold1D}
In this section, we present numerical solutions of the 1D system \eqref{eq:1Dgeom}--\eqref{eq:1DBCs}. We show that, in the absence of morphology, saturating threshold-dependent mechanical feedback, modelled by the following law:
\begin{align}
\frac{\dot{\gamma}}{\gamma} = W(s) + \phi\tanh(n - n^*)\mathrm{H}(n^* - n),\label{eq:1Dsaturatingmechanicalfeedback}
\end{align}
does not produce significantly different behaviours from the linear threshold-dependent mechanical feedback, which is modelled by the growth law \eqref{eq:thresholdbasedmechanicalfeedback}:
\begin{align*}
\frac{\dot{\gamma}}{\gamma} = W(s) + \phi(n - n^*)\mathrm{H}(n^* - n).
\end{align*}

As in Section \ref{sec:formofmechanicalfeedback}, we perform a sweep over the foundation stiffness, $k$, and the strength of mechanical feedback, $\phi$. We classify the resulting incremental growth profiles, $\dot{\gamma}\gamma^{-1}$, according to three different cases: primarily Wnt-driven, where $\dot{\gamma}\gamma^{-1} \approx W(s)$; non-monotonic from the base, where $\dot{\gamma}\gamma^{-1}$ is maximal at $s = 0$, but does not generate the desired proliferative structure; and maximal away from the base and top, where $\dot{\gamma}\gamma^{-1}$ resembles the proliferative structure of the crypt.  The remaining model parameters are set to $\rho = 10$, $\mathcal{S} = 1$, $\sigma_W = 0.24$, and $n^* = -0.4$. The respective phase diagrams for linear and saturating threshold-dependent mechanical feedback are plotted in Fig.\ \ref{fig:1Dlinearvssaturatingthresholdbasedfeedback}. Over the considered values of $k$ and $\phi$, the results are indistinguishable. Not only does the classification of the profiles of $\dot{\gamma}\gamma^{-1}$ split into the same distinct parameter regions, but the representative profiles of $\dot{\gamma}\gamma^{-1}$ are identical. Therefore, in a 1D geometry, linear threshold-dependent mechanical feedback is sufficient to generate the crypt's proliferative structure. However, in 2D, the stretchability $\mathcal{S} \ll 1$ for a realistic crypt geometry, results in a much wider range of stress values than those observed in 1D. Consequently, the effect of mechanical feedback for a linear threshold-dependent law is amplified significantly. To render the computation of homeostatic solutions more amenable to numerical solution, we adopt the 2D equivalent of the growth law \eqref{eq:1Dsaturatingmechanicalfeedback}, where the 1D stress, $n$, is replaced by the axial stress, $n_\tau$, in \eqref{eq:saturatingmechanicalfeedback}. This law is arguably a more physically realistic mechanism for mechanical feedback, as it is likely that the effect of mechanical feedback does not increase indefinitely.

\section{Stability analysis}
\label{app:stabilityanalysis}
We describe the technical details of the procedure that we use to compute the dynamical stability of the homeostatic solutions, given by the set of time-independent solutions, $\mathbf{U}$, and the time-dependent solutions, $\hat{\gamma}(s, t)$ and $\hat{S}_0(s, t)$. However, we note that we do not need to explicitly consider the stability of $\hat{S}_0(s, t)$, as it is embedded in the definition of the flow velocity, $\hat{v}(s)$, through \eqref{eq:flowvelocity}. 

As stated in Sect.\ \ref{sec:stability}, for each time-independent solution $\hat{\nu}(s) \in \mathbf{U}$, we expand the variables as follows:
\begin{align}
\hat{\nu}(s) &= \hat{\nu}^{(0)}(s) + \varepsilon\hat{\nu}^{(1)}(s)\mathrm{e}^{\sigma t},\label{eq:timeindeptsolnpertapp}\\
\hat{\gamma}(s,t) &= \hat{\gamma}_0\hat{\Gamma}^{(0)}(s)\mathrm{e}^{\beta t} + \varepsilon\hat{\gamma}_0\hat{\Gamma}^{(1)}(s)\mathrm{e}^{\beta t}\mathrm{e}^{\sigma t},
\end{align}
where $\beta = g^{(0)}(0)$, the homeostatic incremental growth at the crypt base. At $O(1)$, we recover the homeostatic system outlined in Section \ref{sec:homeostasisconditions}. At $O(\varepsilon)$, we obtain the following linearised boundary value problem:
\begin{align}
&\hat{x}^{(1)'} = -\hat{\theta}^{(1)}\sin\hat{\theta}^{(0)},\label{eq:xeps}\\
&\hat{y}^{(1)'} = \hat{\theta}^{(1)}\cos\hat{\theta}^{(0)},\label{eq:yeps}\\
&\hat{n}_x^{(1)'} = k(\hat{x}^{(1)} - \hat{p}^{(1)}_x),\label{eq:nxeps}\\
&\hat{n}_y^{(1)'} = k(\hat{y}^{(1)} - \hat{p}^{(1)}_y),\label{eq:nyeps}\\
&\hat{p}_x^{(1)'} = \frac{(\sigma +\rho)\hat{p}^{(1)}_x - \rho\hat{x}^{(1)}}{\hat{v}^{(0)}}\nonumber
\\&\qquad\quad+ \frac{\rho(\hat{x}^{(0)} - \hat{p}_x^{(0)})}{(\hat{v}^{(0)})^2}\hat{v}^{(1)},\label{eq:pxeps}\\
&\hat{p}_y^{(1)'} = \frac{(\sigma + \rho)\hat{p}^{(1)}_y -\rho\hat{y}^{(1)}}{\hat{v}^{(0)}}\nonumber
\\&\qquad\quad + \frac{\rho(\hat{y}^{(0)} - \hat{p}_y^{(0)})}{(\hat{v}^{(0)})^2}\hat{v}^{(1)},\label{eq:pyeps}\\
&\hat{\theta}^{(1)'} = \frac{\hat{m}^{(1)}}{\hat{\alpha}^{(0)}} - \frac{\hat{m}^{(0)}}{\left(\hat{\alpha}^{(0)}\right)^2}\hat{\alpha}^{(1)},\label{eq:thetaeps}\\
&\hat{m}^{(1)'} = \hat{n}_x^{(1)}\sin\theta^{(0)} - \hat{n}_y^{(1)}\cos\theta^{(0)}\nonumber
\\&\qquad\quad + \theta^{(1)}(\hat{n}_x^{(0)}\cos\theta^{(0)} + \hat{n}_y^{(0)}\sin\theta^{(0)}),\label{eq:meps}\\
&\hat{v}^{(1)'} = \frac{\hat{\alpha}^{(0)'}}{\alpha^{(0)}}\hat{v}^{(1)} + \frac{\hat{v}^{(0)}}{\hat{\alpha}^{(0)}}\hat{\alpha}^{(1)'}- \phi f'\left(\hat{n}^{(0)}_\tau - \hat{n}^*_\tau\right)\hat{n}^{(1)}_\tau\nonumber
\\&\qquad\quad- \frac{\left(\hat{\alpha}^{(0)'}\hat{v}^{(0)} + \sigma\hat{\alpha}^{(0)}\right)}{\left(\hat{\alpha}^{(0)}\right)^2}\hat{\alpha}^{(1)},\label{eq:veps}\\
&\Gamma^{(1)'} = \frac{(\beta + \sigma - W(s))}{\hat{v}^{(0)}}\Gamma^{(1)} - \frac{\Gamma^{(0)'}}{\hat{v}^{(0)}}\hat{v}^{(1)}\nonumber
\\&\qquad\quad- \frac{\phi f'\left(\hat{n}^{(0)}_\tau - \hat{n}^*_\tau\right)\Gamma^{(0)}}{\hat{v}^{(0)}}\hat{n}^{(1)}_\tau,\label{eq:gammaeps}
\end{align}
where $\hat{\alpha}^{(1)}$ is found by expanding the constitutive relation \eqref{eq:eulerianelasticstretch}:
\begin{align}
\hat{\alpha}^{(1)} &= \mathcal{S}\Big(\hat{n}_x^{(1)}\cos\hat{\theta}^{(0)} + \hat{n}_y^{(1)}\sin\hat{\theta}^{(0)}\nonumber
\\&\qquad\quad + \hat{\theta}^{(1)}\left(\hat{n}_y^{(0)}\cos\hat{\theta}^{(0)} - \hat{n}_x^{(0)}\sin\hat{\theta}^{(0)}\right)\Big).\label{eq:alphaeps}
\end{align}
Consequently, the derivative $\hat{\alpha}^{(1)'}$ can be determined.

The linearised boundary conditions are found by applying the Eulerian boundary conditions \eqref{eq:eulerianBCs} and the velocity boundary condition $\hat{v}(0) = 0$, yielding the left boundary conditions:
\begin{align}
&\hat{x}^{(1)}(0) = 0,\label{eq:leftbcsstart}\\
&\hat{n}_y^{(1)}(0) = 0,\\
&\hat{\theta}^{(1)}(0) = 0,\\
&\hat{v}^{(1)}(0) = 0,\\
&\hat{p}_x^{(1)}(0) = 0,\\
&\hat{p}_y^{(1)}(0)= \frac{\rho}{\sigma + \rho}\hat{y}^{(1)}(0),\\
& \Gamma^{(1)}(0) = \frac{\phi f'(\hat{n}_x^{(0)}(0) - \hat{n}_\tau^*)}{\sigma}\hat{n}_x^{(1)}(0),\label{eq:leftbcsend}
\end{align}
and the right boundary conditions:
\begin{align}
&\hat{x}^{(1)}(l) = 0,\label{eq:rightbcsstart}\\
&\hat{y}^{(1)}(l) = 0,\\
&\hat{\theta}^{(1)}(l) = 0.\label{eq:rightbcsend}
\end{align}
We note that the linearised boundary value problem can be expressed in the compact form, $\hat{\mathbf{x}}_s^{(1)} = \mathbf{A}\hat{\mathbf{x}}^{(1)}$, where $\hat{\mathbf{x}}^{(1)}$ is the linearised solution vector
\begin{align}
\hat{\mathbf{x}}^{(1)} =\left(\hat{x}^{(1)},\hat{y}^{(1)},\hat{n}_x^{(1)},\hat{n}^{(1)}_y,\hat{p}^{(1)}_x,\hat{p}^{(1)}_y,\hat{\theta}^{(1)},\hat{m}^{(1)},\hat{v}^{(1)},\Gamma^{(1)}\right)^T\nonumber,
\end{align}
and $\mathbf{A}$ is the coefficient matrix and a function of both $s$ and $\sigma$. Furthermore, we observe from \eqref{eq:leftbcsstart}--\eqref{eq:leftbcsend} that there are no known left boundary conditions for $\hat{y}^{(1)}$, $\hat{n}_x^{(1)}$ and $\hat{m}^{(1)}$. To overcome this, we write $\hat{\mathbf{x}}^{(1)}$ as three linearly-independent copies:
\begin{align}
\hat{\mathbf{x}}^{(1)} = c_1\hat{\mathbf{x}}_1^{(1)} + c_2\hat{\mathbf{x}}_2^{(1)} + c_3\hat{\mathbf{x}}_3^{(1)},
\end{align}
where $\hat{\mathbf{x}}_1^{(1)}$, $\hat{\mathbf{x}}_2^{(1)}$, and $\hat{\mathbf{x}}_3^{(1)}$ satisfy the following left boundary conditions, respectively:
\begin{align}
&\hat{\mathbf{x}}_1^{(1)}(0) =\left(0,1,0,0,0, \frac{\rho}{\sigma + \rho}, 0, 0, 0, 0\right)^T,\nonumber\\
&\hat{\mathbf{x}}_2^{(1)}(0) =\left(0,0,1,0,0, 0, 0, 0, 0, \frac{\phi f'(\hat{n}_x^{(0)}(0) - \hat{n}_\tau^*)}{\sigma}\right)^T,\nonumber\\
&\hat{\mathbf{x}}_2^{(1)}(0) =\left(0,0,0,0,0, 0, 0, 1, 0, 0\right)^T.
\end{align}
In other words, the left boundary conditions for $\hat{\mathbf{x}}_1^{(1)}(s)$, $\hat{\mathbf{x}}_2^{(1)}(s)$, and $\hat{\mathbf{x}}_3^{(1)}(s)$ correspond to prescribing $\hat{y}^{(1)}(0) = 1$, $\hat{n}_x^{(1)}(0) = 1$, and $\hat{m}^{(1)}(0) = 1$, respectively. By linear independence, the solution for each copy can be obtained independently, as functions of the eigenvalue, $\sigma$. The constants $c_1$ and $c_2$ and $c_3$ are determined by imposing that the right boundary conditions \eqref{eq:rightbcsstart}--\eqref{eq:rightbcsend} are satisfied. This leads to the following matrix equation at $s = l$:
\begin{align}
\left(\begin{array}{ccc}\hat{x}_1^{(1)} & \hat{x}_2^{(1)} & \hat{x}_3^{(1)} \\
				\hat{y}_1^{(1)} & \hat{y}_2^{(1)} & \hat{y}_3^{(1)} \\ 
				\hat{\theta}_1^{(1)} & \hat{\theta}_2^{(1)} & \hat{\theta}_3^{(1)} \end{array}\right)\left(\begin{array}{c}c_1\\c_2\\c_3\end{array}\right) = \left(\begin{array}{c}0\\0\\0\end{array}\right), \quad \mbox{ at }\quad s = l.\label{eq:c1c2c3condition}
\end{align}
A solution to \eqref{eq:c1c2c3condition} exists if and only if the determinant of the matrix on the left-hand side is zero, for a given value of $\sigma$. If all values of $\sigma$ that cause the determinant of \eqref{eq:c1c2c3condition} to vanish satisfy $\mathrm{Re}(\sigma) < 0$, then the dynamic perturbations decay to zero as $t \to \infty$. Therefore, the homeostatic solutions are dynamically stable. However, if there is at least one positive value of $\sigma$, then the dynamic perturbations grow exponentially in time and the homeostatic solutions are dynamically unstable. 
\end{document}